\documentclass[aps,prl,reprint,superscriptaddress,amsmath,amssymb]{revtex4-2}
\usepackage{graphicx}
\usepackage{longtable}
\usepackage{xcolor}
\usepackage[bookmarks,bookmarksnumbered,colorlinks,allcolors=blue]{hyperref}
\usepackage{bookmark}
\usepackage{ifthen}
\usepackage{ulem}
\usepackage{soul}

\newcommand{\ket}[1]{|{#1}\rangle}

\newcommand{\Phiext}[1]{\Phi_{#1}}
\newcommand{\hatsigma}[1]{\hat{\sigma}_{#1}}

\begin{document}
\title{Native approach to controlled-Z gates in inductively coupled fluxonium qubits}

\author{Xizheng Ma}
\thanks{These authors contributed equally to this work}
\affiliation{DAMO Quantum Laboratory, Alibaba Group, Hangzhou, Zhejiang 311121, China}
\author{Gengyan Zhang}
\thanks{These authors contributed equally to this work}
\author{Feng Wu}
\thanks{These authors contributed equally to this work}
\author{Feng Bao}
\author{Xu Chang}
\author{Jianjun Chen}
\thanks{Current address: Xinxiao Electronics Inc., Hangzhou, China}
\author{Hao Deng}
\author{Ran Gao}
\affiliation{DAMO Quantum Laboratory, Alibaba Group, Hangzhou, Zhejiang 311121, China}
\author{Xun Gao}
\affiliation{DAMO Quantum Laboratory, Alibaba Group USA, Bellevue, WA 98004, USA}
\author{Lijuan Hu}
\affiliation{DAMO Quantum Laboratory, Alibaba Group, Hangzhou, Zhejiang 311121, China}
\author{Honghong Ji}
\author{Hsiang-Sheng Ku}
\thanks{Current address: IQM Quantum Computers, 80992 Munich, Germany}
\author{Kannan Lu}
\author{Lu Ma}
\author{Liyong Mao}
\author{Zhijun Song}
\author{Hantao Sun}
\author{Chengchun Tang}
\author{Fei Wang}
\author{Hongcheng Wang}
\author{Tenghui Wang}
\author{Tian Xia}
\author{Make Ying}
\author{Huijuan Zhan}
\author{Tao Zhou}
\author{Mengyu Zhu}
\author{Qingbin Zhu}
\affiliation{DAMO Quantum Laboratory, Alibaba Group, Hangzhou, Zhejiang 311121, China}
\author{Yaoyun Shi}
\affiliation{DAMO Quantum Laboratory, Alibaba Group USA, Bellevue, WA 98004, USA}
\author{Hui-Hai Zhao}
\email{huihai.zhh@alibaba-inc.com}
\affiliation{DAMO Quantum Laboratory, Alibaba Group, Beijing 100102, China}
\author{Chunqing Deng}
\email{chunqing.cd@alibaba-inc.com}
\affiliation{DAMO Quantum Laboratory, Alibaba Group, Hangzhou, Zhejiang 311121, China}

\begin{abstract}
    The fluxonium qubits have emerged as a promising platform for gate-based quantum information processing. 
    However, their extraordinary protection against charge fluctuations comes at a cost: when coupled capacitively, the qubit-qubit interactions are restricted to \textit{XX}-interactions.
    Consequently, effective \textit{ZZ}- or \textit{XZ}-interactions are only constructed either by temporarily populating higher-energy states, or by exploiting perturbative effects under microwave driving.
    Instead, we propose and demonstrate an inductive coupling scheme, which offers a wide selection of native qubit-qubit interactions for fluxonium.
    In particular, we leverage a built-in, flux-controlled \textit{ZZ}-interaction to perform qubit entanglement.
    To combat the increased flux-noise-induced dephasing away from the flux-insensitive position, we use a continuous version of the dynamical decoupling scheme to perform noise filtering. 
    Combining these, we demonstrate a 20~ns controlled-Z (CZ) gate with a mean fidelity of 99.53\%.
    More than confirming the efficacy of our gate scheme, this high-fidelity result also reveals a promising but rarely explored parameter space uniquely suitable for gate operations between fluxonium qubits.
\end{abstract}

\maketitle

Repeated demonstrations of long coherence times~\cite{pop2014,Zhang2021,Somoroff2023} and high-fidelity gate-operations~\cite{Bao2022,Dogan2023,Moskalenko2022,Ding2023}
over the years have firmly established the fluxonium qubits as a promising platform for gate-based quantum computation. Compared to transmons, fluxonium qubits have two obvious advantages:
their low transition energies between the ground and first excited state allows for better coherence times by reducing the effect of dielectric loss~\cite{Lin2018, Zhang2021, Sun2023}; and their large anharmonicity offers a broad spectral range for qubit operations without leakage to higher (noncomputational) excited states~\cite{Dogan2023,Bao2022}. Most experiments, therefore, have treated fluxonium as an improved version of transmon by opting for a capacitive coupling scheme standard for charge-type qubits~\cite{Koch2007}. However, the fluxonium's complete lack of energy dispersion in charge basis, while providing an excellent protection against charge-noise-induced dephasing, means that any interaction mediated through charge is restricted to the transverse direction of the qubit.
Although such \textit{XX}-interaction can be used to perform iSWAP-like gates~\cite{Bao2022,Moskalenko2022}, it precludes easy access to a broader range of two-qubit operations such as CZ or CNOT gates.
Instead, an effective \textit{ZZ}-interaction needs to be constructed~\cite{Ficheux2021,Xiong2022,Ding2023,Moskalenko2022,Simakov2023} by temporarily populating the transmon-like higher excited states,
whose short coherence times  ultimately limit the fidelity of the gates. Alternatively, a second-order effect of the driven dynamics can be exploited~\cite{Dogan2023,Nguyen2022} to create an effective \textit{XZ}-coupling, but such schemes have to balance slow gate speed with large unwanted interactions. 

\begin{figure*}
    \centering
    \includegraphics{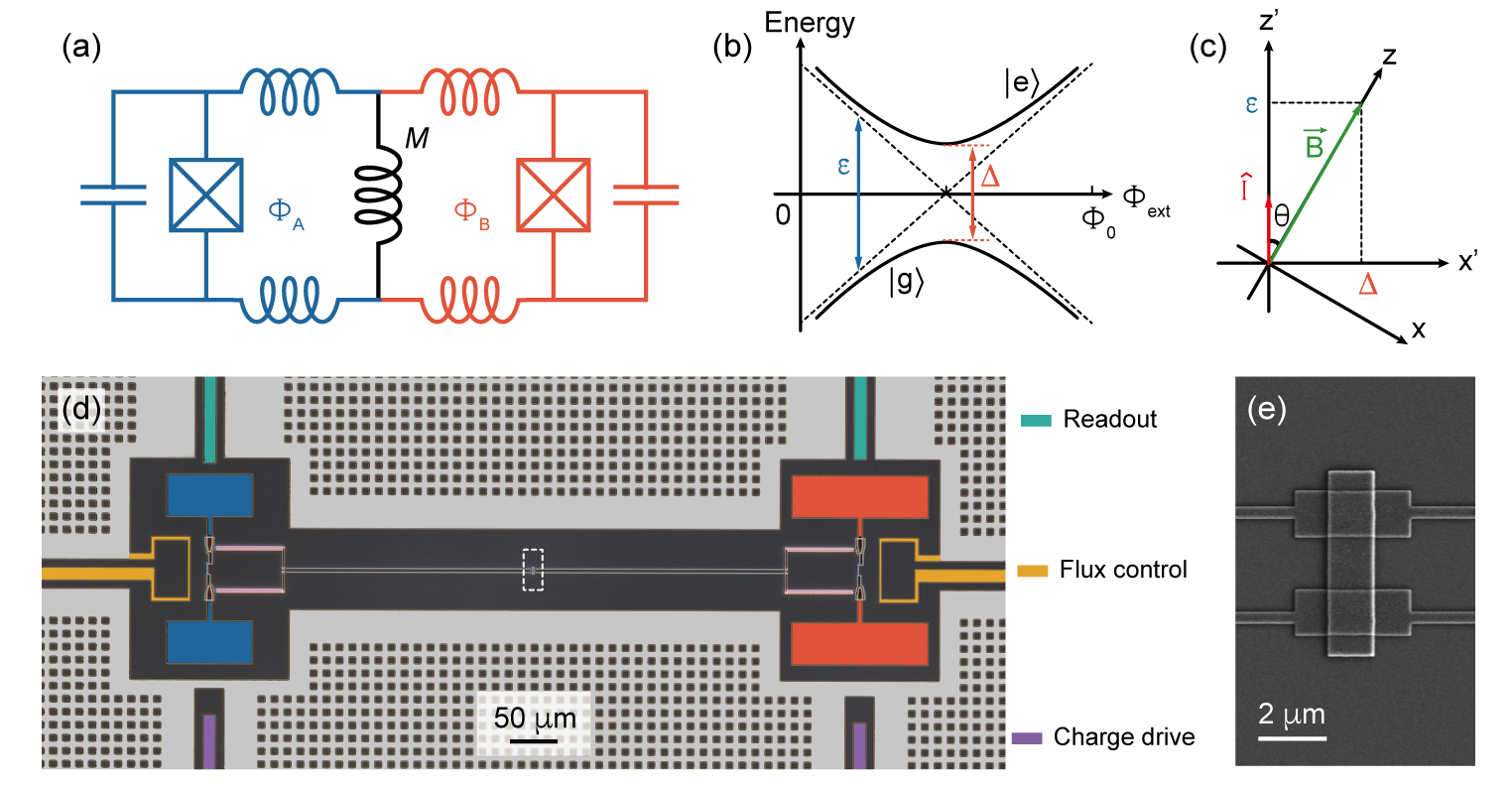}
    \caption{\textbf{Inductively coupled fluxonium pair.} (a) A pair of fluxonium qubits are coupled inductively though a mutual inductance $M$, creating an interaction of the form $M\hat{I}_A\hat{I}_B$, where $i = A$ or $B$ indexes the qubit and $\hat{I}_i$ is the qubit's current operator.
    Each qubit can be individually controlled with an external flux $\Phiext{i}$ applied to its superconducting loop with total inductance $L_i$.
    (b) For each qubit, its ground ($\ket{g}$) and excited ($\ket{e}$) states are superpositions of two persistent current states (dashed lines) with equal currents circulating the loop in opposite directions. Defining the average magnetic energy of the two persistent current states to be zero, their energy difference $\epsilon$ is linearly dependent on $\Phiext{}$ applied to the qubit.
    The degeneracy between these states at $\Phiext{} = \Phi_0/2$ is lifted by the Josephson junction with a tunneling energy $\Delta$.
    (c) Such a system resembles~\cite{Leggett1987} a spin of unit dipole moment subjected to a fictitious magnetic field $\vec{B} = \Delta \hat{x}' + \epsilon \hat{z}'$, where $\hat{z}'$ aligns with $\hat{I}$ (red). By adjusting $\epsilon$ via $\Phiext{}$, we can access a diverse range of interactions by controlling the orientation of the qubit energy quantization axis $\hatsigma{z}$, aligned along the net magnetic field $\hat{z}$ (green) and rotated by an angle $\theta$ with respect to the current operator $\hat{I}$.
    (d) False colored optical image of the inductively coupled fluxonium pair. The white dashed rectangle locates the mutual inductance, created by a galvanic connection that shares two junctions between the two fluxonium loops.
    (e) Scanning electron micrograph showing the two shared overlap~\cite{Wu2017} junctions.}
    \label{fig:Fig1}
\end{figure*}

At its core, fluxonium qubits belong to the family of flux qubits, and are therefore most naturally coupled inductively (\autoref{fig:Fig1}(a)). All flux qubits are essentially superconducting loops interrupted by Josephson junctions. When an external flux $\Phiext{}$ close to half-flux quantum $\Phi_0/2$ is applied to the loop, 
persistent currents $\pm I_p$ circulate the loop in opposite directions, expelling or pulling additional external flux to maintain flux quantization. These persistent current states are coupled via the Josephson junction at a tunneling energy $\Delta$ (\autoref{fig:Fig1}(b)).
By controlling the magnetic energy difference between these states $\epsilon = 2 I_p (\Phiext{} - \Phi_0/2)$, $\Phiext{}$ determines the qubit energy $\hbar \omega_q = \sqrt{\epsilon^2 + \Delta^2}$. More importantly, it also determines the orientation of the current operator $\hat{I}$ flowing across the junction with respect to the energy quantization axis $\hatsigma{z}$ of the qubit (\autoref{fig:Fig1}(c)):
\begin{equation}\label{eqn:spin rotation}
    \hat{I} =  I_p \left(\cos{\theta} \hatsigma{z} - \sin{\theta} \hatsigma{x} \right),
\end{equation}
where $\theta = \text{arctan}(\Delta/\epsilon)$ is the mixing angle between the tunneling energy and the magnetic energy. At $\Phiext{} = \Phi_0/2$, the qubit energy is first order insensitive to $\Phiext{}$, and $\hat{I}$ is perpendicular to the energy quantization axis $\hatsigma{z}$. Away from the degeneracy position, fluctuations in the current not only drives qubit excitation but also alters the qubit energy. Naturally, when a pair of flux qubits are connected via a mutual inductance $M$, circulating currents in one qubit induces current in the other qubit, creating a coupling of the form $M \hat{I}_i \hat{I}_j$, where the subscripts index the qubits.

Indeed, inductive coupling offers a diverse range of entangling interactions via the control of the external flux $\Phiext{}$ applied to the qubits. These interactions are partly investigated by the early proposals~\cite{Mooij1999,Orlando1999,Plourde2004,Rigetti2005,Bertet2006,Niskanen2006,Kerman2008} concerned with connecting the prototypical flux qubits~\cite{Mooij1999}.
However, impeded by their extreme sensitivity to flux noise, these flux qubits saw limited application in gate-based quantum computing and have largely pivoted to the field of quantum annealing~\cite{VanderPloeg2007,Harris2007,Johnson2011,Weber2017}. The few demonstrations~\cite{Niskanen2007,DeGroot2010} of gate-based operations are limited to the flux-degeneracy position, where the lack of first-order flux dispersion alleviates the qubit decoherence but again restricts the qubit interactions to the transverse directions.
By reducing the persistent currents $I_p$ with the increased loop inductance, fluxonium qubits significantly reduce their sensitivity to flux noise~\cite{Paladino2014,Nguyen2019}, paving the way toward high-fidelity entangling operations that require temporary excursions away from the flux-degeneracy position.

\begin{figure*}
    \centering
    \includegraphics{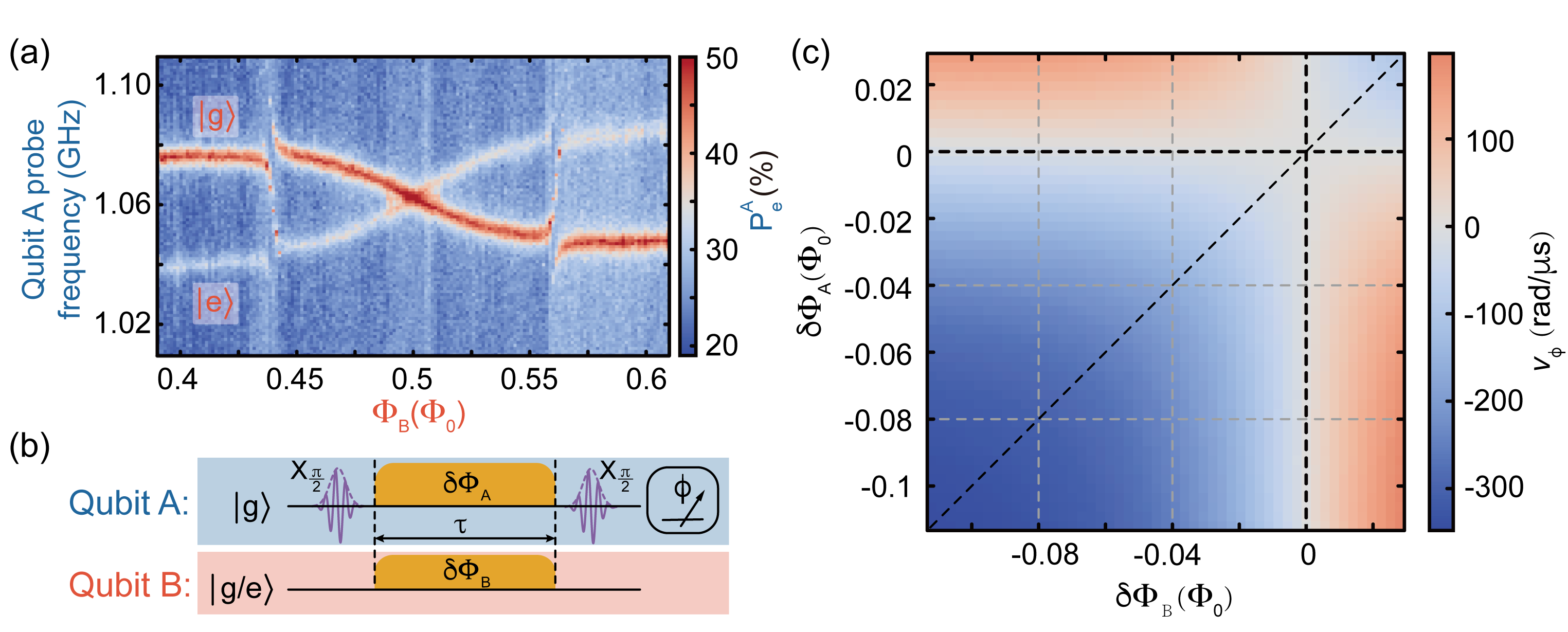}
    \caption{\textbf{A native ZZ-interaction allows for conditional-phase gates.} (a) The spectrum of qubit A is measured as a function of the external flux $\Phiext{B}$ applied to qubit B, while $\Phiext{A} = 0.458~\Phi_0$ is kept as a constant. As a result of the native \textit{ZZ}-interaction, qubit A's resonant frequency is shifted in opposite directions when qubit B is prepared in either $\ket{g}$ or $\ket{e}$. Here, qubit B is prepared in a mixed state with approximately $24\%$ probability in $\ket{e}$, leading to a double-peaked spectrum where the more prominent peak corresponds to qubit B in $\ket{g}$.
    Additionally, level repulsions can be observed when the two qubits come into resonance at $\Phiext{B} = 0.5\pm 0.062~\Phi_0$.
    (b) Using a Ramsey-type experiment, we measure the conditional phase $\phi$ accumulated on qubit A when we bias both qubits away from $\Phi_0/2$ using square pulse of amplitude $\delta \Phi_\text{A,B}$ for duration $\tau$.
    (c) As a function of the pulse amplitudes, we plot the measured accumulation speed of the conditional phase $v_\phi = \phi/\tau$.
    }
    \label{fig:Fig2}
\end{figure*}

In this work, we inductively couple a pair of fluxonium qubits and demonstrate high-fidelity gate-operations by leveraging a native \textit{ZZ}-interaction that is only switched on when both qubits are biased away from their flux-degeneracy positions. 
As shown in \autoref{fig:Fig1}(d,e), the inductive coupling is created by a galvanic connection that shares part of the junction arrays forming the loop inductors.
Under external fluxes $\Phiext{A}$ and $\Phiext{B}$ close to $\Phi_0/2$ (see Supplementary Material), we can write down the general form of the system's Hamiltonian, 
\begin{align}\label{eqn:Hamiltonian}
    H(\Phiext{A},\Phiext{B}) = \frac{\hbar}{2}&\sum_{i = A,B} \omega_i \hatsigma{z}^i + J \sin{\theta_A}\sin{\theta_B} \hatsigma{x}^A\hatsigma{x}^B \nonumber\\
    &- J \sum_{i\neq j} \cos{\theta_i}\sin{\theta_j}\hatsigma{z}^i\hatsigma{x}^j \nonumber\\
    &+ J\cos{\theta_A}\cos{\theta_B} \hatsigma{z}^A\hatsigma{z}^B,
\end{align}
where $J/\hbar = M I_p^A I_p^B/\hbar \approx 2\pi\times 19~\text{MHz}$ is the inductive coupling strength, and $\theta_i = \text{arctan}\left(\Delta_i/\epsilon_i(\Phiext{i})\right)$ is the mixing angle of the $i$-th qubit with $\cos{\theta_i} = 0$ at $\Phiext{i} = \Phi_0/2$. 
In this Hamiltonian, the first line contains the always-on transverse interaction. 
The second line describes an \textit{XZ}-interaction when either qubit is biased away from its flux-degeneracy position. When modulated at an appropriate frequency~\cite{Didier2015,Rigetti2010}, such a native \textit{XZ}-coupling provides a promising path toward fast controlled-NOT gate. 
But for near-static flux modulations considered in this work, qubit $j$ merely acquires an $i$-state-dependent rotation of angle $\left( 2J/\hbar\omega_j\right) \cos{\theta_i}\sin{\theta_j} \hatsigma{z}^i$ that has little consequence for $J \ll \hbar\omega_j$ (see Supplementary Material). Finally, when both qubits are biased away from the half-flux position, the \textit{ZZ}-interaction in the last line of \autoref{eqn:Hamiltonian} induces qubit frequency shifts $\delta \omega_i = 2g_{zz} \langle\hatsigma{z}^j \rangle$ dependent on the state of the other qubit, defining the shorthand $\hbar g_{zz} = J \cos{\theta_A}\cos{\theta_B}$.

To confirm this system Hamiltonian, we perform spectroscopic measurements on qubit A at a constant $\Phiext{A} = 0.458~\Phi_0$ (\autoref{fig:Fig2}(a)) while varying the external flux $\Phiext{B}$ applied to qubit B. When the two qubits come into resonance, the transverse interaction causes a coherent exchange of qubit energy, which manifests as level repulsions at $\Phiext{B} = 0.5 \pm 0.062~\Phi_0$. Away from the resonant positions, this coherent exchange is suppressed by the qubit detuning. Instead, the effect of the \textit{ZZ}-interaction dominates. 
Because qubit B is initialized in a thermal state with approximately 24\% probability of finding $\ket{e}$, the spectrum of qubit A is double-peaked, where the more and less prominent peaks respectively correspond to qubit A's frequency when qubit B is in state $\ket{g}$ and $\ket{e}$. The distance between these peaks therefore directly corresponds to $g_{zz}$, adjustable via the control of the external flux.

Leveraging this flux-controlled \textit{ZZ}-coupling, we implement two-qubit conditional-phase gates by simultaneously applying flux pulses on $\Phiext{A,B}(t)$ to both qubits. Using a Ramsey-type experiment (\autoref{fig:Fig2}(b)), we characterize the accumulation speed of the conditional-phase under flux pulses of length $\tau$,
\begin{equation}\label{eqn:v_phi}
    v_\phi = \frac{1}{\tau}\int_0^\tau 4g_{zz}\left(\Phiext{A}(t),\Phiext{B}(t)\right) dt.
\end{equation}
\autoref{fig:Fig2}(c) shows $v_\phi$ as a function of how far the qubits are biased away from the half-flux position under square pulses of amplitude $\delta \Phi_{A,B}$ and duration $\tau = 200~\text{ns}$. In this experiment, we enveloped the square pulses with a tanh function with a characteristic rise time $t_r = 2~\text{ns}$, slow enough to ensure that the flux modulations do not significantly alter the qubits' excitation via diabatic passages (see Supplementary Material).
Note that when either qubit is kept at the half-flux position ($\delta \Phi_i = 0$), a conditional-phase still accumulates slowly. Indeed, this slow accumulation is the result of a residual \textit{ZZ}-coupling $g_{zz}^\text{res}$, caused by transverse interactions between higher-energy qubit states~\cite{Ficheux2021,Nguyen2022}. Importantly, this residual \textit{ZZ}-coupling does not contribute any two-qubit gate errors because it is naturally absorbed into \autoref{eqn:v_phi} during gate calibration. Meanwhile,  $g_{zz}^\text{res} \approx 2 \pi \times 17.4~\text{kHz}$ at the idle position (see Supplementary Material), where both qubits are parked at $\Phi_0/2$ , contributes a negligible error in the range of $10^{-5}$ to single-qubit operations. We stress that, because we leverage a native \textit{ZZ}-interaction that is first order to the coupling strength $J$, we could easily attain a large on-off ratio in $g_{zz}$ close to $10^3$ without relying on demanding cancellation engineering~\cite{Dogan2023,Xiong2022,Nguyen2022, Moskalenko2022, Ding2023}. Indeed, by simply increasing the distance of both qubits from $\Phi_0/2$, we can drastically increase the accumulation speed of the conditional-phase and perform two-qubit CZ gates ($v_\phi \tau = \pi$) as fast as $9~\text{ns}$. However, this exceptional gate-speed comes at the cost of increased qubit sensitivity to flux noise and reduced coherence times.

\begin{figure}
    \centering
    \includegraphics{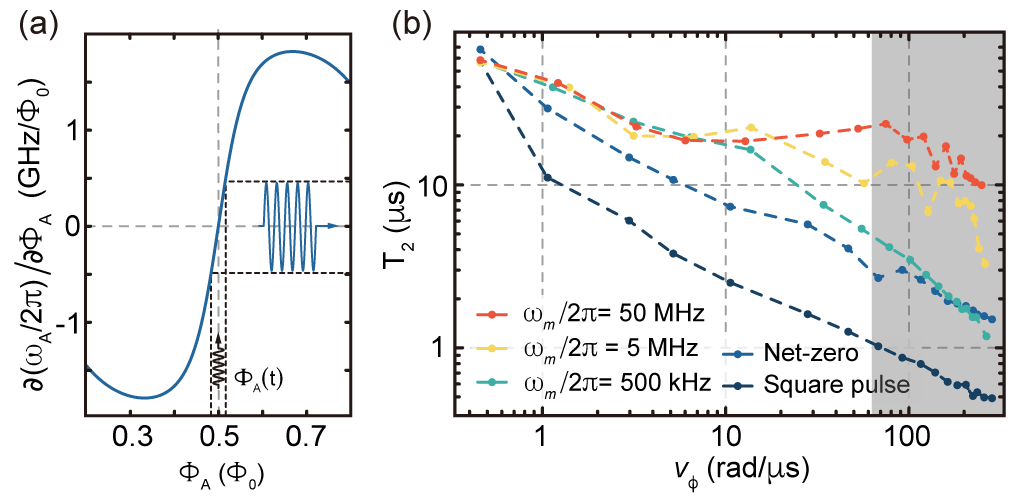}
    \caption{\textbf{Using dynamical decoupling to combat flux-noise induced dephasing.} (a) The first-order derivative of qubit A's frequency dispersion, $\partial \omega_A /\partial \Phiext{A}$, is plotted as a function of $\Phiext{A}$. A sinusoidal flux modulation about $\Phi_0/2$ produces an approximately-sinusoidal modulation of $\partial \omega_A /\partial \Phiext{A}$, which averages to 0 over integer periods.
    (b) We measure the characteristic decoherence times ($T_2$) of Qubit A under different flux modulations $\Phiext{A}(t)$ while Qubit B is kept at $\Phi_0/2$. For easy comparison, the amplitude of the flux modulation is expressed in $v_\phi$ had the same flux modulation been applied to both qubits ($\Phiext{A}(t) = \Phiext{B}(t)$). The shaded region corresponds to $v_\phi$ large enough to support CZ gates under $50~\text{ns}$.
    }
    \label{fig:Fig3}
\end{figure}

To combat the effect of flux noise during two-qubit gate-operations, we embed a continuous version of the dynamical decoupling scheme~\cite{Viola1998} to our flux-control pulses. Specifically, we sinusoidally modulate both flux pulses at frequency $\omega_m$ with equal phase, resulting in a near-sinusoidal modulation on the slope of either qubit's flux dispersion, $\partial \omega_i/\partial \Phiext{i}$, which averages to zero over integer periods (\autoref{fig:Fig3}(a)). Consequently, the qubit dephasing becomes sensitive only to a narrow region of noises with frequencies close to $\omega_m$ (see Supplementary Material). Compared to a net-zero decoupling scheme~\cite{Rol2019,Negirneac2021}, where each flux pulse consists two back-to-back square pulses of equal duration but opposite amplitude, our scheme provides an \textit{in-situ} selection of noise frequency while requiring only microwave controls.
It is important to note that, because $\partial \hbar\omega_i/\partial \Phiext{i} = 2I_p^i\cos{\theta_i}$, this energy dispersion to external flux is precisely the source of the native \textit{ZZ}-interaction we exploit in two-qubit gates: $g_{zz}\propto \left(\partial \omega_A/\partial \Phiext{A}\right)\left(\partial \omega_B/\partial \Phiext{B}\right)$. 
Yet, while our sinusoidal flux modulations average the energy dispersion of either qubit to zero to reduce dephasing, their correlation nevertheless preserves a nonzero \textit{ZZ}-interaction, or $v_\phi$, averaged over integer periods. 
In \autoref{fig:Fig3}(b), we demonstrate the efficacy of our dynamical decoupling scheme by measuring qubit A's characteristic decoherence time $T_2$ under different flux pulses $\Phiext{A}(t)$ while qubit B is kept at $\Phi_0/2$. For easy comparison (see Supplementary Material), the amplitude of $\Phiext{A}(t)$ is expressed in $v_\phi$ had both qubits been simultaneously modulated with the same pulse ($\Phiext{B}(t) = \Phiext{A}(t)$, diagonal dashed-line in \autoref{fig:Fig2}(c)). 
Because the qubit dephasing is dominated by $1/f$-flux noise, a clear improvement in $T_2$ can be observed when the modulation frequency is increased from $500~\text{kHz}$ to $50~\text{MHz}$.
Compared to the case without any dynamical decoupling scheme, the sinusoidal flux modulation improves the coherence time of qubit A by more than an order of magnitude at modulation amplitudes large enough to support fast CZ gates under $50~\text{ns}$ (shaded region).

Employing the sinusoidal dynamical decoupling with $\omega_m = 2\pi \times 50~\text{MHz}$, we calibrate a $20~\text{ns}$ CZ gate. The detailed calibration process can be found in the supplementary material. 
Unfortunately, we suffer from relatively large flux-pulse distortions that require us to append another $20~\text{ns}$ idle time after every CZ gate, effectively doubling our gate time. 

\begin{figure}
    \centering
    \includegraphics{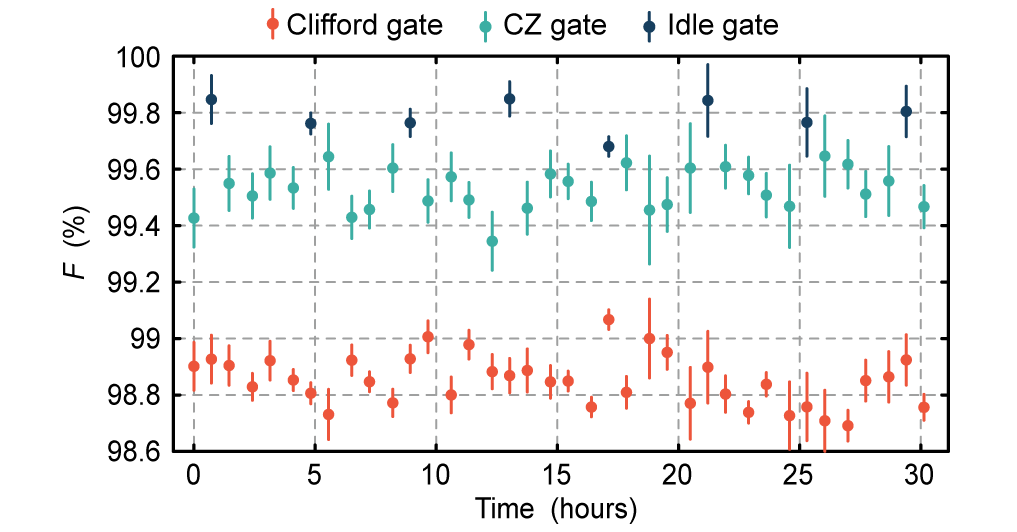}
    \caption{\textbf{Gate performance.} The gate performances are monitored over a period of 30 hours using the standard RB techniques. The error bars correspond to fitting uncertainties.
    }
    \label{fig:Fig4}
\end{figure}

Nevertheless, we demonstrate in \autoref{fig:Fig4} our ability to perform high-fidelity gate-operations using randomized benchmarking (RB)~\cite{Magesan2012,Barends2014}.
Monitored over 30 hours, we measure a mean Clifford fidelity of $\overline{F_{C}} = 98.85 \pm 0.08\%$, and a mean CZ gate fidelity of $\overline{F_\text{CZ}} = 99.53\pm 0.07\%$, where the uncertainty intervals capture the time fluctuations in the measured fidelities. 
Because our device suffers from serious TLS poisoning whose effect fluctuates over time (see Supplementary Material), it is immensely difficult to accurately predict the decoherence limit of our CZ gate. Instead, we provide a sense of the decoherence effect on the gate fidelity by measuring the mean $\overline{{F}_\text{idle}} = 99.78 \pm 0.05\%$ of a $40~\text{ns}$ idle gate, limited by the coherence times at the idle position (see Supplementary Material). In doing so, we find that approximately half of our CZ gate error comes from decoherence sources that also suppress our qubit coherence times at the degeneracy positions to below $100~\mu\text{s}$ (see Supplementary Material), a subpar performance for fluxonium qubits.
Improvements in this performance or reductions in the idle time after each CZ gate therefore could significantly improve our gate fidelities.

More than confirming the efficacy of our gate scheme, our high-fidelity result also reveals a promising but rarely explored parameter space for gate operations.
Whereas transmons traded away rich interactions similar to those described in \autoref{eqn:Hamiltonian} in favor of an improved coherence time, prototypical flux qubits simply suffer too much decoherence to effectively leverage them. 
Because of their reduced sensitivity to flux noise, fluxnoium qubits can be operated away from their degeneracy positions, and are therefore uniquely suited to exploit the diverse interactions enabled by inductive-coupling for gate operations.

In summary, we demonstrated a particular synergy between fluxonium qubits and inductive coupling schemes that leads to a native \textit{ZZ}-interaction when both qubits are biased away from the flux-degeneracy positions.
By adjusting the external flux applied to the qubits, we can tune the \textit{ZZ}-interaction strength over three orders of magnitude, enabling fast entangling operations with minimal adverse effects to single-qubit operations.  
Finally, by embedding a sinusoidal dynamical decoupling scheme into the control sequences, we suppressed the additional dephasing introduced by the two-qubit operations and demonstrated high-fidelity CZ gates that does not involve higher energy states.
Looking forward, tunable couplers based on inductively coupled SQUIDs or fluxonium qubits~\cite{Weiss2022} may be a prerequisite for performing the quantum operations demonstrated in this work at a much larger scale. 
Alternatively, the thus far neglected native \textit{XZ}-interactions may also hold a promising path toward highly-scalable~\cite{Nguyen2022} entangling operations.

\begin{acknowledgments}
We thank the broader DAMO Quantum Laboratory team for technical support. We also thank Xin Wan for insightful discussions.
\end{acknowledgments}

\bibliographystyle{apsrev4-2}
\bibliography{ref}

\begin{thebibliography}{32}%
\makeatletter
\providecommand \@ifxundefined [1]{%
 \@ifx{#1\undefined}
}%
\providecommand \@ifnum [1]{%
 \ifnum #1\expandafter \@firstoftwo
 \else \expandafter \@secondoftwo
 \fi
}%
\providecommand \@ifx [1]{%
 \ifx #1\expandafter \@firstoftwo
 \else \expandafter \@secondoftwo
 \fi
}%
\providecommand \natexlab [1]{#1}%
\providecommand \enquote  [1]{``#1''}%
\providecommand \bibnamefont  [1]{#1}%
\providecommand \bibfnamefont [1]{#1}%
\providecommand \citenamefont [1]{#1}%
\providecommand \href@noop [0]{\@secondoftwo}%
\providecommand \href [0]{\begingroup \@sanitize@url \@href}%
\providecommand \@href[1]{\@@startlink{#1}\@@href}%
\providecommand \@@href[1]{\endgroup#1\@@endlink}%
\providecommand \@sanitize@url [0]{\catcode `\\12\catcode `\$12\catcode
  `\&12\catcode `\#12\catcode `\^12\catcode `\_12\catcode `\%12\relax}%
\providecommand \@@startlink[1]{}%
\providecommand \@@endlink[0]{}%
\providecommand \url  [0]{\begingroup\@sanitize@url \@url }%
\providecommand \@url [1]{\endgroup\@href {#1}{\urlprefix }}%
\providecommand \urlprefix  [0]{URL }%
\providecommand \Eprint [0]{\href }%
\providecommand \doibase [0]{https://doi.org/}%
\providecommand \selectlanguage [0]{\@gobble}%
\providecommand \bibinfo  [0]{\@secondoftwo}%
\providecommand \bibfield  [0]{\@secondoftwo}%
\providecommand \translation [1]{[#1]}%
\providecommand \BibitemOpen [0]{}%
\providecommand \bibitemStop [0]{}%
\providecommand \bibitemNoStop [0]{.\EOS\space}%
\providecommand \EOS [0]{\spacefactor3000\relax}%
\providecommand \BibitemShut  [1]{\csname bibitem#1\endcsname}%
\let\auto@bib@innerbib\@empty
\bibitem [{\citenamefont {Pop}\ \emph {et~al.}(2014)\citenamefont {Pop},
  \citenamefont {Geerlings}, \citenamefont {Catelani}, \citenamefont
  {Schoelkopf}, \citenamefont {Glazman},\ and\ \citenamefont
  {Devoret}}]{pop2014}%
  \BibitemOpen
  \bibfield  {author} {\bibinfo {author} {\bibfnamefont {I.~M.}\ \bibnamefont
  {Pop}}, \bibinfo {author} {\bibfnamefont {K.}~\bibnamefont {Geerlings}},
  \bibinfo {author} {\bibfnamefont {G.}~\bibnamefont {Catelani}}, \bibinfo
  {author} {\bibfnamefont {R.~J.}\ \bibnamefont {Schoelkopf}}, \bibinfo
  {author} {\bibfnamefont {L.~I.}\ \bibnamefont {Glazman}},\ and\ \bibinfo
  {author} {\bibfnamefont {M.~H.}\ \bibnamefont {Devoret}},\ }\href
  {https://doi.org/10.1038/nature13017} {\bibfield  {journal} {\bibinfo
  {journal} {Nature}\ }\textbf {\bibinfo {volume} {508}},\ \bibinfo {pages}
  {369} (\bibinfo {year} {2014})}\BibitemShut {NoStop}%
\bibitem [{\citenamefont {Gustavsson}\ \emph {et~al.}(2016)\citenamefont
  {Gustavsson}, \citenamefont {Yan}, \citenamefont {Catelani}, \citenamefont
  {Bylander}, \citenamefont {Kamal}, \citenamefont {Birenbaum}, \citenamefont
  {Hover}, \citenamefont {Rosenberg}, \citenamefont {Samach}, \citenamefont
  {Sears}, \citenamefont {Weber}, \citenamefont {Yoder}, \citenamefont
  {Clarke}, \citenamefont {Kerman}, \citenamefont {Yoshihara}, \citenamefont
  {Nakamura}, \citenamefont {Orlando},\ and\ \citenamefont
  {Oliver}}]{Gustavsson2016}%
  \BibitemOpen
  \bibfield  {author} {\bibinfo {author} {\bibfnamefont {S.}~\bibnamefont
  {Gustavsson}}, \bibinfo {author} {\bibfnamefont {F.}~\bibnamefont {Yan}},
  \bibinfo {author} {\bibfnamefont {G.}~\bibnamefont {Catelani}}, \bibinfo
  {author} {\bibfnamefont {J.}~\bibnamefont {Bylander}}, \bibinfo {author}
  {\bibfnamefont {A.}~\bibnamefont {Kamal}}, \bibinfo {author} {\bibfnamefont
  {J.}~\bibnamefont {Birenbaum}}, \bibinfo {author} {\bibfnamefont
  {D.}~\bibnamefont {Hover}}, \bibinfo {author} {\bibfnamefont
  {D.}~\bibnamefont {Rosenberg}}, \bibinfo {author} {\bibfnamefont
  {G.}~\bibnamefont {Samach}}, \bibinfo {author} {\bibfnamefont {A.~P.}\
  \bibnamefont {Sears}}, \bibinfo {author} {\bibfnamefont {S.~J.}\ \bibnamefont
  {Weber}}, \bibinfo {author} {\bibfnamefont {J.~L.}\ \bibnamefont {Yoder}},
  \bibinfo {author} {\bibfnamefont {J.}~\bibnamefont {Clarke}}, \bibinfo
  {author} {\bibfnamefont {A.~J.}\ \bibnamefont {Kerman}}, \bibinfo {author}
  {\bibfnamefont {F.}~\bibnamefont {Yoshihara}}, \bibinfo {author}
  {\bibfnamefont {Y.}~\bibnamefont {Nakamura}}, \bibinfo {author}
  {\bibfnamefont {T.~P.}\ \bibnamefont {Orlando}},\ and\ \bibinfo {author}
  {\bibfnamefont {W.~D.}\ \bibnamefont {Oliver}},\ }\href
  {https://doi.org/10.1126/science.aah5844} {\bibfield  {journal} {\bibinfo
  {journal} {Science}\ }\textbf {\bibinfo {volume} {354}},\ \bibinfo {pages}
  {1573} (\bibinfo {year} {2016})}\BibitemShut {NoStop}%
\bibitem [{\citenamefont {Bao}\ \emph {et~al.}(2022)\citenamefont {Bao},
  \citenamefont {Deng}, \citenamefont {Ding}, \citenamefont {Gao},
  \citenamefont {Gao}, \citenamefont {Huang}, \citenamefont {Jiang},
  \citenamefont {Ku}, \citenamefont {Li}, \citenamefont {Ma}, \citenamefont
  {Ni}, \citenamefont {Qin}, \citenamefont {Song}, \citenamefont {Sun},
  \citenamefont {Tang}, \citenamefont {Wang}, \citenamefont {Wu}, \citenamefont
  {Xia}, \citenamefont {Yu}, \citenamefont {Zhang}, \citenamefont {Zhang},
  \citenamefont {Zhang}, \citenamefont {Zhou}, \citenamefont {Zhu},
  \citenamefont {Shi}, \citenamefont {Chen}, \citenamefont {Zhao},\ and\
  \citenamefont {Deng}}]{Bao2022}%
  \BibitemOpen
  \bibfield  {author} {\bibinfo {author} {\bibfnamefont {F.}~\bibnamefont
  {Bao}}, \bibinfo {author} {\bibfnamefont {H.}~\bibnamefont {Deng}}, \bibinfo
  {author} {\bibfnamefont {D.}~\bibnamefont {Ding}}, \bibinfo {author}
  {\bibfnamefont {R.}~\bibnamefont {Gao}}, \bibinfo {author} {\bibfnamefont
  {X.}~\bibnamefont {Gao}}, \bibinfo {author} {\bibfnamefont {C.}~\bibnamefont
  {Huang}}, \bibinfo {author} {\bibfnamefont {X.}~\bibnamefont {Jiang}},
  \bibinfo {author} {\bibfnamefont {H.-S.}\ \bibnamefont {Ku}}, \bibinfo
  {author} {\bibfnamefont {Z.}~\bibnamefont {Li}}, \bibinfo {author}
  {\bibfnamefont {X.}~\bibnamefont {Ma}}, \bibinfo {author} {\bibfnamefont
  {X.}~\bibnamefont {Ni}}, \bibinfo {author} {\bibfnamefont {J.}~\bibnamefont
  {Qin}}, \bibinfo {author} {\bibfnamefont {Z.}~\bibnamefont {Song}}, \bibinfo
  {author} {\bibfnamefont {H.}~\bibnamefont {Sun}}, \bibinfo {author}
  {\bibfnamefont {C.}~\bibnamefont {Tang}}, \bibinfo {author} {\bibfnamefont
  {T.}~\bibnamefont {Wang}}, \bibinfo {author} {\bibfnamefont {F.}~\bibnamefont
  {Wu}}, \bibinfo {author} {\bibfnamefont {T.}~\bibnamefont {Xia}}, \bibinfo
  {author} {\bibfnamefont {W.}~\bibnamefont {Yu}}, \bibinfo {author}
  {\bibfnamefont {F.}~\bibnamefont {Zhang}}, \bibinfo {author} {\bibfnamefont
  {G.}~\bibnamefont {Zhang}}, \bibinfo {author} {\bibfnamefont
  {X.}~\bibnamefont {Zhang}}, \bibinfo {author} {\bibfnamefont
  {J.}~\bibnamefont {Zhou}}, \bibinfo {author} {\bibfnamefont {X.}~\bibnamefont
  {Zhu}}, \bibinfo {author} {\bibfnamefont {Y.}~\bibnamefont {Shi}}, \bibinfo
  {author} {\bibfnamefont {J.}~\bibnamefont {Chen}}, \bibinfo {author}
  {\bibfnamefont {H.-H.}\ \bibnamefont {Zhao}},\ and\ \bibinfo {author}
  {\bibfnamefont {C.}~\bibnamefont {Deng}},\ }\href
  {https://doi.org/10.1103/PhysRevLett.129.010502} {\bibfield  {journal}
  {\bibinfo  {journal} {Phys. Rev. Lett.}\ }\textbf {\bibinfo {volume} {129}},\
  \bibinfo {pages} {010502} (\bibinfo {year} {2022})}\BibitemShut {NoStop}%
\bibitem [{\citenamefont {Zhang}\ \emph {et~al.}(2021)\citenamefont {Zhang},
  \citenamefont {Chakram}, \citenamefont {Roy}, \citenamefont {Earnest},
  \citenamefont {Lu}, \citenamefont {Huang}, \citenamefont {Weiss},
  \citenamefont {Koch},\ and\ \citenamefont {Schuster}}]{Zhang2021}%
  \BibitemOpen
  \bibfield  {author} {\bibinfo {author} {\bibfnamefont {H.}~\bibnamefont
  {Zhang}}, \bibinfo {author} {\bibfnamefont {S.}~\bibnamefont {Chakram}},
  \bibinfo {author} {\bibfnamefont {T.}~\bibnamefont {Roy}}, \bibinfo {author}
  {\bibfnamefont {N.}~\bibnamefont {Earnest}}, \bibinfo {author} {\bibfnamefont
  {Y.}~\bibnamefont {Lu}}, \bibinfo {author} {\bibfnamefont {Z.}~\bibnamefont
  {Huang}}, \bibinfo {author} {\bibfnamefont {D.~K.}\ \bibnamefont {Weiss}},
  \bibinfo {author} {\bibfnamefont {J.}~\bibnamefont {Koch}},\ and\ \bibinfo
  {author} {\bibfnamefont {D.~I.}\ \bibnamefont {Schuster}},\ }\href
  {https://doi.org/10.1103/PhysRevX.11.011010} {\bibfield  {journal} {\bibinfo
  {journal} {Phys. Rev. X}\ }\textbf {\bibinfo {volume} {11}},\ \bibinfo
  {pages} {011010} (\bibinfo {year} {2021})}\BibitemShut {NoStop}%
\bibitem [{\citenamefont {Somoroff}\ \emph {et~al.}(2023)\citenamefont
  {Somoroff}, \citenamefont {Ficheux}, \citenamefont {Mencia}, \citenamefont
  {Xiong}, \citenamefont {Kuzmin},\ and\ \citenamefont
  {Manucharyan}}]{Somoroff2023}%
  \BibitemOpen
  \bibfield  {author} {\bibinfo {author} {\bibfnamefont {A.}~\bibnamefont
  {Somoroff}}, \bibinfo {author} {\bibfnamefont {Q.}~\bibnamefont {Ficheux}},
  \bibinfo {author} {\bibfnamefont {R.~A.}\ \bibnamefont {Mencia}}, \bibinfo
  {author} {\bibfnamefont {H.}~\bibnamefont {Xiong}}, \bibinfo {author}
  {\bibfnamefont {R.}~\bibnamefont {Kuzmin}},\ and\ \bibinfo {author}
  {\bibfnamefont {V.~E.}\ \bibnamefont {Manucharyan}},\ }\href
  {https://doi.org/10.1103/PhysRevLett.130.267001} {\bibfield  {journal}
  {\bibinfo  {journal} {Phys. Rev. Lett.}\ }\textbf {\bibinfo {volume} {130}},\
  \bibinfo {pages} {267001} (\bibinfo {year} {2023})}\BibitemShut {NoStop}%
\bibitem [{\citenamefont {Barends}\ \emph {et~al.}(2013)\citenamefont
  {Barends}, \citenamefont {Kelly}, \citenamefont {Megrant}, \citenamefont
  {Sank}, \citenamefont {Jeffrey}, \citenamefont {Chen}, \citenamefont {Yin},
  \citenamefont {Chiaro}, \citenamefont {Mutus}, \citenamefont {Neill},
  \citenamefont {O'Malley}, \citenamefont {Roushan}, \citenamefont {Wenner},
  \citenamefont {White}, \citenamefont {Cleland},\ and\ \citenamefont
  {Martinis}}]{Barends2013}%
  \BibitemOpen
  \bibfield  {author} {\bibinfo {author} {\bibfnamefont {R.}~\bibnamefont
  {Barends}}, \bibinfo {author} {\bibfnamefont {J.}~\bibnamefont {Kelly}},
  \bibinfo {author} {\bibfnamefont {A.}~\bibnamefont {Megrant}}, \bibinfo
  {author} {\bibfnamefont {D.}~\bibnamefont {Sank}}, \bibinfo {author}
  {\bibfnamefont {E.}~\bibnamefont {Jeffrey}}, \bibinfo {author} {\bibfnamefont
  {Y.}~\bibnamefont {Chen}}, \bibinfo {author} {\bibfnamefont {Y.}~\bibnamefont
  {Yin}}, \bibinfo {author} {\bibfnamefont {B.}~\bibnamefont {Chiaro}},
  \bibinfo {author} {\bibfnamefont {J.}~\bibnamefont {Mutus}}, \bibinfo
  {author} {\bibfnamefont {C.}~\bibnamefont {Neill}}, \bibinfo {author}
  {\bibfnamefont {P.}~\bibnamefont {O'Malley}}, \bibinfo {author}
  {\bibfnamefont {P.}~\bibnamefont {Roushan}}, \bibinfo {author} {\bibfnamefont
  {J.}~\bibnamefont {Wenner}}, \bibinfo {author} {\bibfnamefont {T.~C.}\
  \bibnamefont {White}}, \bibinfo {author} {\bibfnamefont {A.~N.}\ \bibnamefont
  {Cleland}},\ and\ \bibinfo {author} {\bibfnamefont {J.~M.}\ \bibnamefont
  {Martinis}},\ }\href {https://doi.org/10.1103/PhysRevLett.111.080502}
  {\bibfield  {journal} {\bibinfo  {journal} {Phys. Rev. Lett.}\ }\textbf
  {\bibinfo {volume} {111}},\ \bibinfo {pages} {080502} (\bibinfo {year}
  {2013})}\BibitemShut {NoStop}%
\bibitem [{\citenamefont {Klimov}\ \emph {et~al.}(2018)\citenamefont {Klimov},
  \citenamefont {Kelly}, \citenamefont {Chen}, \citenamefont {Neeley},
  \citenamefont {Megrant}, \citenamefont {Burkett}, \citenamefont {Barends},
  \citenamefont {Arya}, \citenamefont {Chiaro}, \citenamefont {Chen},
  \citenamefont {Dunsworth}, \citenamefont {Fowler}, \citenamefont {Foxen},
  \citenamefont {Gidney}, \citenamefont {Giustina}, \citenamefont {Graff},
  \citenamefont {Huang}, \citenamefont {Jeffrey}, \citenamefont {Lucero},
  \citenamefont {Mutus}, \citenamefont {Naaman}, \citenamefont {Neill},
  \citenamefont {Quintana}, \citenamefont {Roushan}, \citenamefont {Sank},
  \citenamefont {Vainsencher}, \citenamefont {Wenner}, \citenamefont {White},
  \citenamefont {Boixo}, \citenamefont {Babbush}, \citenamefont {Smelyanskiy},
  \citenamefont {Neven},\ and\ \citenamefont {Martinis}}]{Klimov2018}%
  \BibitemOpen
  \bibfield  {author} {\bibinfo {author} {\bibfnamefont {P.~V.}\ \bibnamefont
  {Klimov}}, \bibinfo {author} {\bibfnamefont {J.}~\bibnamefont {Kelly}},
  \bibinfo {author} {\bibfnamefont {Z.}~\bibnamefont {Chen}}, \bibinfo {author}
  {\bibfnamefont {M.}~\bibnamefont {Neeley}}, \bibinfo {author} {\bibfnamefont
  {A.}~\bibnamefont {Megrant}}, \bibinfo {author} {\bibfnamefont
  {B.}~\bibnamefont {Burkett}}, \bibinfo {author} {\bibfnamefont
  {R.}~\bibnamefont {Barends}}, \bibinfo {author} {\bibfnamefont
  {K.}~\bibnamefont {Arya}}, \bibinfo {author} {\bibfnamefont {B.}~\bibnamefont
  {Chiaro}}, \bibinfo {author} {\bibfnamefont {Y.}~\bibnamefont {Chen}},
  \bibinfo {author} {\bibfnamefont {A.}~\bibnamefont {Dunsworth}}, \bibinfo
  {author} {\bibfnamefont {A.}~\bibnamefont {Fowler}}, \bibinfo {author}
  {\bibfnamefont {B.}~\bibnamefont {Foxen}}, \bibinfo {author} {\bibfnamefont
  {C.}~\bibnamefont {Gidney}}, \bibinfo {author} {\bibfnamefont
  {M.}~\bibnamefont {Giustina}}, \bibinfo {author} {\bibfnamefont
  {R.}~\bibnamefont {Graff}}, \bibinfo {author} {\bibfnamefont
  {T.}~\bibnamefont {Huang}}, \bibinfo {author} {\bibfnamefont
  {E.}~\bibnamefont {Jeffrey}}, \bibinfo {author} {\bibfnamefont
  {E.}~\bibnamefont {Lucero}}, \bibinfo {author} {\bibfnamefont {J.~Y.}\
  \bibnamefont {Mutus}}, \bibinfo {author} {\bibfnamefont {O.}~\bibnamefont
  {Naaman}}, \bibinfo {author} {\bibfnamefont {C.}~\bibnamefont {Neill}},
  \bibinfo {author} {\bibfnamefont {C.}~\bibnamefont {Quintana}}, \bibinfo
  {author} {\bibfnamefont {P.}~\bibnamefont {Roushan}}, \bibinfo {author}
  {\bibfnamefont {D.}~\bibnamefont {Sank}}, \bibinfo {author} {\bibfnamefont
  {A.}~\bibnamefont {Vainsencher}}, \bibinfo {author} {\bibfnamefont
  {J.}~\bibnamefont {Wenner}}, \bibinfo {author} {\bibfnamefont {T.~C.}\
  \bibnamefont {White}}, \bibinfo {author} {\bibfnamefont {S.}~\bibnamefont
  {Boixo}}, \bibinfo {author} {\bibfnamefont {R.}~\bibnamefont {Babbush}},
  \bibinfo {author} {\bibfnamefont {V.~N.}\ \bibnamefont {Smelyanskiy}},
  \bibinfo {author} {\bibfnamefont {H.}~\bibnamefont {Neven}},\ and\ \bibinfo
  {author} {\bibfnamefont {J.~M.}\ \bibnamefont {Martinis}},\ }\href
  {https://doi.org/10.1103/PhysRevLett.121.090502} {\bibfield  {journal}
  {\bibinfo  {journal} {Phys. Rev. Lett.}\ }\textbf {\bibinfo {volume} {121}},\
  \bibinfo {pages} {090502} (\bibinfo {year} {2018})}\BibitemShut {NoStop}%
\bibitem [{\citenamefont {Ding}\ \emph {et~al.}(2021)\citenamefont {Ding},
  \citenamefont {Ku}, \citenamefont {Shi},\ and\ \citenamefont
  {Zhao}}]{Ding2021}%
  \BibitemOpen
  \bibfield  {author} {\bibinfo {author} {\bibfnamefont {D.}~\bibnamefont
  {Ding}}, \bibinfo {author} {\bibfnamefont {H.-S.}\ \bibnamefont {Ku}},
  \bibinfo {author} {\bibfnamefont {Y.}~\bibnamefont {Shi}},\ and\ \bibinfo
  {author} {\bibfnamefont {H.-H.}\ \bibnamefont {Zhao}},\ }\href
  {https://doi.org/10.1103/PhysRevB.103.174501} {\bibfield  {journal} {\bibinfo
   {journal} {Phys. Rev. B}\ }\textbf {\bibinfo {volume} {103}},\ \bibinfo
  {pages} {174501} (\bibinfo {year} {2021})}\BibitemShut {NoStop}%
\bibitem [{\citenamefont {Zhang}\ \emph {et~al.}(2017)\citenamefont {Zhang},
  \citenamefont {Liu}, \citenamefont {Raftery},\ and\ \citenamefont
  {Houck}}]{Zhang2017}%
  \BibitemOpen
  \bibfield  {author} {\bibinfo {author} {\bibfnamefont {G.}~\bibnamefont
  {Zhang}}, \bibinfo {author} {\bibfnamefont {Y.}~\bibnamefont {Liu}}, \bibinfo
  {author} {\bibfnamefont {J.~J.}\ \bibnamefont {Raftery}},\ and\ \bibinfo
  {author} {\bibfnamefont {A.~A.}\ \bibnamefont {Houck}},\ }\href
  {https://doi.org/10.1038/s41534-016-0002-2} {\bibfield  {journal} {\bibinfo
  {journal} {npj Quantum Inf}\ }\textbf {\bibinfo {volume} {3}},\ \bibinfo
  {pages} {1} (\bibinfo {year} {2017})}\BibitemShut {NoStop}%
\bibitem [{\citenamefont {Rigetti}\ and\ \citenamefont
  {Devoret}(2010)}]{Rigetti2010}%
  \BibitemOpen
  \bibfield  {author} {\bibinfo {author} {\bibfnamefont {C.}~\bibnamefont
  {Rigetti}}\ and\ \bibinfo {author} {\bibfnamefont {M.}~\bibnamefont
  {Devoret}},\ }\href {https://doi.org/10.1103/PhysRevB.81.134507} {\bibfield
  {journal} {\bibinfo  {journal} {Phys. Rev. B}\ }\textbf {\bibinfo {volume}
  {81}},\ \bibinfo {pages} {1} (\bibinfo {year} {2010})}\BibitemShut {NoStop}%
\bibitem [{\citenamefont {Didier}\ \emph {et~al.}(2015)\citenamefont {Didier},
  \citenamefont {Bourassa},\ and\ \citenamefont {Blais}}]{Didier2015}%
  \BibitemOpen
  \bibfield  {author} {\bibinfo {author} {\bibfnamefont {N.}~\bibnamefont
  {Didier}}, \bibinfo {author} {\bibfnamefont {J.}~\bibnamefont {Bourassa}},\
  and\ \bibinfo {author} {\bibfnamefont {A.}~\bibnamefont {Blais}},\ }\href
  {https://doi.org/10.1103/PhysRevLett.115.203601} {\bibfield  {journal}
  {\bibinfo  {journal} {Phys. Rev. Lett.}\ }\textbf {\bibinfo {volume} {115}},\
  \bibinfo {pages} {203601} (\bibinfo {year} {2015})}\BibitemShut {NoStop}%
\bibitem [{\citenamefont {Amin}(2009)}]{Amin2009}%
  \BibitemOpen
  \bibfield  {author} {\bibinfo {author} {\bibfnamefont {M.~H.~S.}\
  \bibnamefont {Amin}},\ }\href
  {https://doi.org/10.1103/PhysRevLett.102.220401} {\bibfield  {journal}
  {\bibinfo  {journal} {Phys. Rev. Lett.}\ }\textbf {\bibinfo {volume} {102}},\
  \bibinfo {pages} {220401} (\bibinfo {year} {2009})}\BibitemShut {NoStop}%
\bibitem [{\citenamefont {Bertet}\ \emph {et~al.}(2006)\citenamefont {Bertet},
  \citenamefont {Harmans},\ and\ \citenamefont {Mooij}}]{Bertet2006}%
  \BibitemOpen
  \bibfield  {author} {\bibinfo {author} {\bibfnamefont {P.}~\bibnamefont
  {Bertet}}, \bibinfo {author} {\bibfnamefont {C.~J.}\ \bibnamefont
  {Harmans}},\ and\ \bibinfo {author} {\bibfnamefont {J.~E.}\ \bibnamefont
  {Mooij}},\ }\href {https://doi.org/10.1103/PhysRevB.73.064512} {\bibfield
  {journal} {\bibinfo  {journal} {Phys. Rev. B}\ }\textbf {\bibinfo {volume}
  {73}},\ \bibinfo {pages} {1} (\bibinfo {year} {2006})}\BibitemShut {NoStop}%
\bibitem [{\citenamefont {Niskanen}\ \emph {et~al.}(2006)\citenamefont
  {Niskanen}, \citenamefont {Nakamura},\ and\ \citenamefont
  {Tsai}}]{Niskanen2006}%
  \BibitemOpen
  \bibfield  {author} {\bibinfo {author} {\bibfnamefont {A.~O.}\ \bibnamefont
  {Niskanen}}, \bibinfo {author} {\bibfnamefont {Y.}~\bibnamefont {Nakamura}},\
  and\ \bibinfo {author} {\bibfnamefont {J.~S.}\ \bibnamefont {Tsai}},\ }\href
  {https://doi.org/10.1103/PhysRevB.73.094506} {\bibfield  {journal} {\bibinfo
  {journal} {Phys. Rev. B}\ }\textbf {\bibinfo {volume} {73}},\ \bibinfo
  {pages} {1} (\bibinfo {year} {2006})}\BibitemShut {NoStop}%
\bibitem [{\citenamefont {Niskanen}\ \emph {et~al.}(2007)\citenamefont
  {Niskanen}, \citenamefont {Harrabi}, \citenamefont {Yoshihara}, \citenamefont
  {Nakamura}, \citenamefont {Lloyd},\ and\ \citenamefont
  {Tsai}}]{Niskanen2007}%
  \BibitemOpen
  \bibfield  {author} {\bibinfo {author} {\bibfnamefont {A.~O.}\ \bibnamefont
  {Niskanen}}, \bibinfo {author} {\bibfnamefont {K.}~\bibnamefont {Harrabi}},
  \bibinfo {author} {\bibfnamefont {F.}~\bibnamefont {Yoshihara}}, \bibinfo
  {author} {\bibfnamefont {Y.}~\bibnamefont {Nakamura}}, \bibinfo {author}
  {\bibfnamefont {S.}~\bibnamefont {Lloyd}},\ and\ \bibinfo {author}
  {\bibfnamefont {J.~S.}\ \bibnamefont {Tsai}},\ }\href
  {https://doi.org/10.1126/science.1141324} {\bibfield  {journal} {\bibinfo
  {journal} {Science}\ }\textbf {\bibinfo {volume} {316}},\ \bibinfo {pages}
  {723} (\bibinfo {year} {2007})}\BibitemShut {NoStop}%
\bibitem [{\citenamefont {Weiss}\ \emph {et~al.}(2022)\citenamefont {Weiss},
  \citenamefont {Zhang}, \citenamefont {Ding}, \citenamefont {Ma},
  \citenamefont {Schuster},\ and\ \citenamefont {Koch}}]{Weiss2022}%
  \BibitemOpen
  \bibfield  {author} {\bibinfo {author} {\bibfnamefont {D.}~\bibnamefont
  {Weiss}}, \bibinfo {author} {\bibfnamefont {H.}~\bibnamefont {Zhang}},
  \bibinfo {author} {\bibfnamefont {C.}~\bibnamefont {Ding}}, \bibinfo {author}
  {\bibfnamefont {Y.}~\bibnamefont {Ma}}, \bibinfo {author} {\bibfnamefont
  {D.~I.}\ \bibnamefont {Schuster}},\ and\ \bibinfo {author} {\bibfnamefont
  {J.}~\bibnamefont {Koch}},\ }\href
  {https://doi.org/10.1103/PRXQuantum.3.040336} {\bibfield  {journal} {\bibinfo
   {journal} {PRX Quantum}\ }\textbf {\bibinfo {volume} {3}},\ \bibinfo {pages}
  {040336} (\bibinfo {year} {2022})}\BibitemShut {NoStop}%
\bibitem [{\citenamefont {Ficheux}\ \emph {et~al.}(2021)\citenamefont
  {Ficheux}, \citenamefont {Nguyen}, \citenamefont {Somoroff}, \citenamefont
  {Xiong}, \citenamefont {Nesterov}, \citenamefont {Vavilov},\ and\
  \citenamefont {Manucharyan}}]{Ficheux2021}%
  \BibitemOpen
  \bibfield  {author} {\bibinfo {author} {\bibfnamefont {Q.}~\bibnamefont
  {Ficheux}}, \bibinfo {author} {\bibfnamefont {L.~B.}\ \bibnamefont {Nguyen}},
  \bibinfo {author} {\bibfnamefont {A.}~\bibnamefont {Somoroff}}, \bibinfo
  {author} {\bibfnamefont {H.}~\bibnamefont {Xiong}}, \bibinfo {author}
  {\bibfnamefont {K.~N.}\ \bibnamefont {Nesterov}}, \bibinfo {author}
  {\bibfnamefont {M.~G.}\ \bibnamefont {Vavilov}},\ and\ \bibinfo {author}
  {\bibfnamefont {V.~E.}\ \bibnamefont {Manucharyan}},\ }\href
  {https://doi.org/10.1103/PhysRevX.11.021026} {\bibfield  {journal} {\bibinfo
  {journal} {Phys. Rev. X}\ }\textbf {\bibinfo {volume} {11}},\ \bibinfo
  {pages} {021026} (\bibinfo {year} {2021})}\BibitemShut {NoStop}%
\bibitem [{\citenamefont {Nguyen}\ \emph {et~al.}(2022)\citenamefont {Nguyen},
  \citenamefont {Koolstra}, \citenamefont {Kim}, \citenamefont {Morvan},
  \citenamefont {Chistolini}, \citenamefont {Singh}, \citenamefont {Nesterov},
  \citenamefont {J{\"{u}}nger}, \citenamefont {Chen}, \citenamefont
  {Pedramrazi}, \citenamefont {Mitchell}, \citenamefont {Kreikebaum},
  \citenamefont {Puri}, \citenamefont {Santiago},\ and\ \citenamefont
  {Siddiqi}}]{Nguyen2022}%
  \BibitemOpen
  \bibfield  {author} {\bibinfo {author} {\bibfnamefont {L.~B.}\ \bibnamefont
  {Nguyen}}, \bibinfo {author} {\bibfnamefont {G.}~\bibnamefont {Koolstra}},
  \bibinfo {author} {\bibfnamefont {Y.}~\bibnamefont {Kim}}, \bibinfo {author}
  {\bibfnamefont {A.}~\bibnamefont {Morvan}}, \bibinfo {author} {\bibfnamefont
  {T.}~\bibnamefont {Chistolini}}, \bibinfo {author} {\bibfnamefont
  {S.}~\bibnamefont {Singh}}, \bibinfo {author} {\bibfnamefont {K.~N.}\
  \bibnamefont {Nesterov}}, \bibinfo {author} {\bibfnamefont {C.}~\bibnamefont
  {J{\"{u}}nger}}, \bibinfo {author} {\bibfnamefont {L.}~\bibnamefont {Chen}},
  \bibinfo {author} {\bibfnamefont {Z.}~\bibnamefont {Pedramrazi}}, \bibinfo
  {author} {\bibfnamefont {B.~K.}\ \bibnamefont {Mitchell}}, \bibinfo {author}
  {\bibfnamefont {J.~M.}\ \bibnamefont {Kreikebaum}}, \bibinfo {author}
  {\bibfnamefont {S.}~\bibnamefont {Puri}}, \bibinfo {author} {\bibfnamefont
  {D.~I.}\ \bibnamefont {Santiago}},\ and\ \bibinfo {author} {\bibfnamefont
  {I.}~\bibnamefont {Siddiqi}},\ }\href
  {https://doi.org/10.1103/PRXQuantum.3.037001} {\bibfield  {journal} {\bibinfo
   {journal} {PRX Quantum}\ }\textbf {\bibinfo {volume} {3}},\ \bibinfo {pages}
  {037001} (\bibinfo {year} {2022})}\BibitemShut {NoStop}%
\bibitem [{\citenamefont {Magesan}\ \emph {et~al.}(2012)\citenamefont
  {Magesan}, \citenamefont {Gambetta}, \citenamefont {Johnson}, \citenamefont
  {Ryan}, \citenamefont {Chow}, \citenamefont {Merkel}, \citenamefont
  {da~Silva}, \citenamefont {Keefe}, \citenamefont {Rothwell}, \citenamefont
  {Ohki}, \citenamefont {Ketchen},\ and\ \citenamefont
  {Steffen}}]{Magesan2012}%
  \BibitemOpen
  \bibfield  {author} {\bibinfo {author} {\bibfnamefont {E.}~\bibnamefont
  {Magesan}}, \bibinfo {author} {\bibfnamefont {J.~M.}\ \bibnamefont
  {Gambetta}}, \bibinfo {author} {\bibfnamefont {B.~R.}\ \bibnamefont
  {Johnson}}, \bibinfo {author} {\bibfnamefont {C.~A.}\ \bibnamefont {Ryan}},
  \bibinfo {author} {\bibfnamefont {J.~M.}\ \bibnamefont {Chow}}, \bibinfo
  {author} {\bibfnamefont {S.~T.}\ \bibnamefont {Merkel}}, \bibinfo {author}
  {\bibfnamefont {M.~P.}\ \bibnamefont {da~Silva}}, \bibinfo {author}
  {\bibfnamefont {G.~A.}\ \bibnamefont {Keefe}}, \bibinfo {author}
  {\bibfnamefont {M.~B.}\ \bibnamefont {Rothwell}}, \bibinfo {author}
  {\bibfnamefont {T.~A.}\ \bibnamefont {Ohki}}, \bibinfo {author}
  {\bibfnamefont {M.~B.}\ \bibnamefont {Ketchen}},\ and\ \bibinfo {author}
  {\bibfnamefont {M.}~\bibnamefont {Steffen}},\ }\href
  {https://doi.org/10.1103/PhysRevLett.109.080505} {\bibfield  {journal}
  {\bibinfo  {journal} {Phys. Rev. Lett.}\ }\textbf {\bibinfo {volume} {109}},\
  \bibinfo {pages} {080505} (\bibinfo {year} {2012})}\BibitemShut {NoStop}%
\bibitem [{\citenamefont {Barends}\ \emph {et~al.}(2014)\citenamefont
  {Barends}, \citenamefont {Kelly}, \citenamefont {Megrant}, \citenamefont
  {Veitia}, \citenamefont {Sank}, \citenamefont {Jeffrey}, \citenamefont
  {White}, \citenamefont {Mutus}, \citenamefont {Fowler}, \citenamefont
  {Campbell}, \citenamefont {Chen}, \citenamefont {Chen}, \citenamefont
  {Chiaro}, \citenamefont {Dunsworth}, \citenamefont {Neill}, \citenamefont
  {O'Malley}, \citenamefont {Roushan}, \citenamefont {Vainsencher},
  \citenamefont {Wenner}, \citenamefont {Korotkov}, \citenamefont {Cleland},\
  and\ \citenamefont {Martinis}}]{Barends2014}%
  \BibitemOpen
  \bibfield  {author} {\bibinfo {author} {\bibfnamefont {R.}~\bibnamefont
  {Barends}}, \bibinfo {author} {\bibfnamefont {J.}~\bibnamefont {Kelly}},
  \bibinfo {author} {\bibfnamefont {A.}~\bibnamefont {Megrant}}, \bibinfo
  {author} {\bibfnamefont {A.}~\bibnamefont {Veitia}}, \bibinfo {author}
  {\bibfnamefont {D.}~\bibnamefont {Sank}}, \bibinfo {author} {\bibfnamefont
  {E.}~\bibnamefont {Jeffrey}}, \bibinfo {author} {\bibfnamefont {T.~C.}\
  \bibnamefont {White}}, \bibinfo {author} {\bibfnamefont {J.}~\bibnamefont
  {Mutus}}, \bibinfo {author} {\bibfnamefont {A.~G.}\ \bibnamefont {Fowler}},
  \bibinfo {author} {\bibfnamefont {B.}~\bibnamefont {Campbell}}, \bibinfo
  {author} {\bibfnamefont {Y.}~\bibnamefont {Chen}}, \bibinfo {author}
  {\bibfnamefont {Z.}~\bibnamefont {Chen}}, \bibinfo {author} {\bibfnamefont
  {B.}~\bibnamefont {Chiaro}}, \bibinfo {author} {\bibfnamefont
  {A.}~\bibnamefont {Dunsworth}}, \bibinfo {author} {\bibfnamefont
  {C.}~\bibnamefont {Neill}}, \bibinfo {author} {\bibfnamefont
  {P.}~\bibnamefont {O'Malley}}, \bibinfo {author} {\bibfnamefont
  {P.}~\bibnamefont {Roushan}}, \bibinfo {author} {\bibfnamefont
  {A.}~\bibnamefont {Vainsencher}}, \bibinfo {author} {\bibfnamefont
  {J.}~\bibnamefont {Wenner}}, \bibinfo {author} {\bibfnamefont {A.~N.}\
  \bibnamefont {Korotkov}}, \bibinfo {author} {\bibfnamefont {A.~N.}\
  \bibnamefont {Cleland}},\ and\ \bibinfo {author} {\bibfnamefont {J.~M.}\
  \bibnamefont {Martinis}},\ }\href {https://doi.org/10.1038/nature13171}
  {\bibfield  {journal} {\bibinfo  {journal} {Nature}\ }\textbf {\bibinfo
  {volume} {508}},\ \bibinfo {pages} {500} (\bibinfo {year}
  {2014})}\BibitemShut {NoStop}%
\bibitem [{\citenamefont {Ithier}\ \emph {et~al.}(2005)\citenamefont {Ithier},
  \citenamefont {Collin}, \citenamefont {Joyez}, \citenamefont {Meeson},
  \citenamefont {Vion}, \citenamefont {Esteve}, \citenamefont {Chiarello},
  \citenamefont {Shnirman}, \citenamefont {Makhlin}, \citenamefont {Schriefl},\
  and\ \citenamefont {Sch{\"{o}}n}}]{Ithier2005}%
  \BibitemOpen
  \bibfield  {author} {\bibinfo {author} {\bibfnamefont {G.}~\bibnamefont
  {Ithier}}, \bibinfo {author} {\bibfnamefont {E.}~\bibnamefont {Collin}},
  \bibinfo {author} {\bibfnamefont {P.}~\bibnamefont {Joyez}}, \bibinfo
  {author} {\bibfnamefont {P.~J.}\ \bibnamefont {Meeson}}, \bibinfo {author}
  {\bibfnamefont {D.}~\bibnamefont {Vion}}, \bibinfo {author} {\bibfnamefont
  {D.}~\bibnamefont {Esteve}}, \bibinfo {author} {\bibfnamefont
  {F.}~\bibnamefont {Chiarello}}, \bibinfo {author} {\bibfnamefont
  {A.}~\bibnamefont {Shnirman}}, \bibinfo {author} {\bibfnamefont
  {Y.}~\bibnamefont {Makhlin}}, \bibinfo {author} {\bibfnamefont
  {J.}~\bibnamefont {Schriefl}},\ and\ \bibinfo {author} {\bibfnamefont
  {G.}~\bibnamefont {Sch{\"{o}}n}},\ }\href
  {https://doi.org/10.1103/PhysRevB.72.134519} {\bibfield  {journal} {\bibinfo
  {journal} {Phys. Rev. B}\ }\textbf {\bibinfo {volume} {72}},\ \bibinfo
  {pages} {134519} (\bibinfo {year} {2005})}\BibitemShut {NoStop}%
\bibitem [{\citenamefont {Bylander}\ \emph {et~al.}(2011)\citenamefont
  {Bylander}, \citenamefont {Gustavsson}, \citenamefont {Yan}, \citenamefont
  {Yoshihara}, \citenamefont {Harrabi}, \citenamefont {Fitch}, \citenamefont
  {Cory}, \citenamefont {Nakamura}, \citenamefont {Tsai},\ and\ \citenamefont
  {Oliver}}]{Bylander2011}%
  \BibitemOpen
  \bibfield  {author} {\bibinfo {author} {\bibfnamefont {J.}~\bibnamefont
  {Bylander}}, \bibinfo {author} {\bibfnamefont {S.}~\bibnamefont
  {Gustavsson}}, \bibinfo {author} {\bibfnamefont {F.}~\bibnamefont {Yan}},
  \bibinfo {author} {\bibfnamefont {F.}~\bibnamefont {Yoshihara}}, \bibinfo
  {author} {\bibfnamefont {K.}~\bibnamefont {Harrabi}}, \bibinfo {author}
  {\bibfnamefont {G.}~\bibnamefont {Fitch}}, \bibinfo {author} {\bibfnamefont
  {D.~G.}\ \bibnamefont {Cory}}, \bibinfo {author} {\bibfnamefont
  {Y.}~\bibnamefont {Nakamura}}, \bibinfo {author} {\bibfnamefont {J.-s.}\
  \bibnamefont {Tsai}},\ and\ \bibinfo {author} {\bibfnamefont {W.~D.}\
  \bibnamefont {Oliver}},\ }\href {https://doi.org/10.1038/nphys1994}
  {\bibfield  {journal} {\bibinfo  {journal} {Nature Phys}\ }\textbf {\bibinfo
  {volume} {7}},\ \bibinfo {pages} {565} (\bibinfo {year} {2011})}\BibitemShut
  {NoStop}%
\bibitem [{\citenamefont {Huang}\ \emph {et~al.}(2021)\citenamefont {Huang},
  \citenamefont {Mundada}, \citenamefont {Gyenis}, \citenamefont {Schuster},
  \citenamefont {Houck},\ and\ \citenamefont {Koch}}]{Huang2021}%
  \BibitemOpen
  \bibfield  {author} {\bibinfo {author} {\bibfnamefont {Z.}~\bibnamefont
  {Huang}}, \bibinfo {author} {\bibfnamefont {P.~S.}\ \bibnamefont {Mundada}},
  \bibinfo {author} {\bibfnamefont {A.}~\bibnamefont {Gyenis}}, \bibinfo
  {author} {\bibfnamefont {D.~I.}\ \bibnamefont {Schuster}}, \bibinfo {author}
  {\bibfnamefont {A.~A.}\ \bibnamefont {Houck}},\ and\ \bibinfo {author}
  {\bibfnamefont {J.}~\bibnamefont {Koch}},\ }\href
  {https://doi.org/10.1103/PhysRevApplied.15.034065} {\bibfield  {journal}
  {\bibinfo  {journal} {Phys. Rev. Appl.}\ }\textbf {\bibinfo {volume} {15}},\
  \bibinfo {pages} {034065} (\bibinfo {year} {2021})}\BibitemShut {NoStop}%
\bibitem [{\citenamefont {Landau}(1932)}]{Landau1932}%
  \BibitemOpen
  \bibfield  {author} {\bibinfo {author} {\bibfnamefont {L.}~\bibnamefont
  {Landau}},\ }\href@noop {} {\bibfield  {journal} {\bibinfo  {journal}
  {Physikalische Zeitschrift der Sowjetunion}\ }\textbf {\bibinfo {volume}
  {2}},\ \bibinfo {pages} {46} (\bibinfo {year} {1932})}\BibitemShut {NoStop}%
\bibitem [{\citenamefont {Zener}(1932)}]{Zener1932}%
  \BibitemOpen
  \bibfield  {author} {\bibinfo {author} {\bibfnamefont {C.}~\bibnamefont
  {Zener}},\ }\href {https://doi.org/10.1098/rspa.1932.0165} {\bibfield
  {journal} {\bibinfo  {journal} {Proc. R. Soc. London A}\ }\textbf {\bibinfo
  {volume} {137}},\ \bibinfo {pages} {696} (\bibinfo {year}
  {1932})}\BibitemShut {NoStop}%
\bibitem [{\citenamefont {Martinis}\ and\ \citenamefont
  {Geller}(2014)}]{Martinis2014}%
  \BibitemOpen
  \bibfield  {author} {\bibinfo {author} {\bibfnamefont {J.~M.}\ \bibnamefont
  {Martinis}}\ and\ \bibinfo {author} {\bibfnamefont {M.~R.}\ \bibnamefont
  {Geller}},\ }\href {https://doi.org/10.1103/PhysRevA.90.022307} {\bibfield
  {journal} {\bibinfo  {journal} {Phys. Rev. A}\ }\textbf {\bibinfo {volume}
  {90}},\ \bibinfo {pages} {022307} (\bibinfo {year} {2014})}\BibitemShut
  {NoStop}%
\bibitem [{\citenamefont {Ghosh}\ \emph {et~al.}(2013)\citenamefont {Ghosh},
  \citenamefont {Galiautdinov}, \citenamefont {Zhou}, \citenamefont {Korotkov},
  \citenamefont {Martinis},\ and\ \citenamefont {Geller}}]{Ghosh2013}%
  \BibitemOpen
  \bibfield  {author} {\bibinfo {author} {\bibfnamefont {J.}~\bibnamefont
  {Ghosh}}, \bibinfo {author} {\bibfnamefont {A.}~\bibnamefont {Galiautdinov}},
  \bibinfo {author} {\bibfnamefont {Z.}~\bibnamefont {Zhou}}, \bibinfo {author}
  {\bibfnamefont {A.~N.}\ \bibnamefont {Korotkov}}, \bibinfo {author}
  {\bibfnamefont {J.~M.}\ \bibnamefont {Martinis}},\ and\ \bibinfo {author}
  {\bibfnamefont {M.~R.}\ \bibnamefont {Geller}},\ }\href
  {https://doi.org/10.1103/PhysRevA.87.022309} {\bibfield  {journal} {\bibinfo
  {journal} {Phys. Rev. A}\ }\textbf {\bibinfo {volume} {87}},\ \bibinfo
  {pages} {022309} (\bibinfo {year} {2013})}\BibitemShut {NoStop}%
\bibitem [{\citenamefont {McKay}\ \emph {et~al.}(2017)\citenamefont {McKay},
  \citenamefont {Wood}, \citenamefont {Sheldon}, \citenamefont {Chow},\ and\
  \citenamefont {Gambetta}}]{McKay2017}%
  \BibitemOpen
  \bibfield  {author} {\bibinfo {author} {\bibfnamefont {D.~C.}\ \bibnamefont
  {McKay}}, \bibinfo {author} {\bibfnamefont {C.~J.}\ \bibnamefont {Wood}},
  \bibinfo {author} {\bibfnamefont {S.}~\bibnamefont {Sheldon}}, \bibinfo
  {author} {\bibfnamefont {J.~M.}\ \bibnamefont {Chow}},\ and\ \bibinfo
  {author} {\bibfnamefont {J.~M.}\ \bibnamefont {Gambetta}},\ }\href
  {https://doi.org/10.1103/PhysRevA.96.022330} {\bibfield  {journal} {\bibinfo
  {journal} {Phys. Rev. A}\ }\textbf {\bibinfo {volume} {96}},\ \bibinfo
  {pages} {022330} (\bibinfo {year} {2017})}\BibitemShut {NoStop}%
\bibitem [{\citenamefont {Kelly}\ \emph {et~al.}(2014)\citenamefont {Kelly},
  \citenamefont {Barends}, \citenamefont {Campbell}, \citenamefont {Chen},
  \citenamefont {Chen}, \citenamefont {Chiaro}, \citenamefont {Dunsworth},
  \citenamefont {Fowler}, \citenamefont {Hoi}, \citenamefont {Jeffrey},
  \citenamefont {Megrant}, \citenamefont {Mutus}, \citenamefont {Neill},
  \citenamefont {O'Malley}, \citenamefont {Quintana}, \citenamefont {Roushan},
  \citenamefont {Sank}, \citenamefont {Vainsencher}, \citenamefont {Wenner},
  \citenamefont {White}, \citenamefont {Cleland},\ and\ \citenamefont
  {Martinis}}]{Kelly2014}%
  \BibitemOpen
  \bibfield  {author} {\bibinfo {author} {\bibfnamefont {J.}~\bibnamefont
  {Kelly}}, \bibinfo {author} {\bibfnamefont {R.}~\bibnamefont {Barends}},
  \bibinfo {author} {\bibfnamefont {B.}~\bibnamefont {Campbell}}, \bibinfo
  {author} {\bibfnamefont {Y.}~\bibnamefont {Chen}}, \bibinfo {author}
  {\bibfnamefont {Z.}~\bibnamefont {Chen}}, \bibinfo {author} {\bibfnamefont
  {B.}~\bibnamefont {Chiaro}}, \bibinfo {author} {\bibfnamefont
  {A.}~\bibnamefont {Dunsworth}}, \bibinfo {author} {\bibfnamefont {A.~G.}\
  \bibnamefont {Fowler}}, \bibinfo {author} {\bibfnamefont {I.-C.}\
  \bibnamefont {Hoi}}, \bibinfo {author} {\bibfnamefont {E.}~\bibnamefont
  {Jeffrey}}, \bibinfo {author} {\bibfnamefont {A.}~\bibnamefont {Megrant}},
  \bibinfo {author} {\bibfnamefont {J.}~\bibnamefont {Mutus}}, \bibinfo
  {author} {\bibfnamefont {C.}~\bibnamefont {Neill}}, \bibinfo {author}
  {\bibfnamefont {P.~J.~J.}\ \bibnamefont {O'Malley}}, \bibinfo {author}
  {\bibfnamefont {C.}~\bibnamefont {Quintana}}, \bibinfo {author}
  {\bibfnamefont {P.}~\bibnamefont {Roushan}}, \bibinfo {author} {\bibfnamefont
  {D.}~\bibnamefont {Sank}}, \bibinfo {author} {\bibfnamefont {A.}~\bibnamefont
  {Vainsencher}}, \bibinfo {author} {\bibfnamefont {J.}~\bibnamefont {Wenner}},
  \bibinfo {author} {\bibfnamefont {T.~C.}\ \bibnamefont {White}}, \bibinfo
  {author} {\bibfnamefont {A.~N.}\ \bibnamefont {Cleland}},\ and\ \bibinfo
  {author} {\bibfnamefont {J.~M.}\ \bibnamefont {Martinis}},\ }\href
  {https://doi.org/10.1103/PhysRevLett.112.240504} {\bibfield  {journal}
  {\bibinfo  {journal} {Phys. Rev. Lett.}\ }\textbf {\bibinfo {volume} {112}},\
  \bibinfo {pages} {240504} (\bibinfo {year} {2014})}\BibitemShut {NoStop}%
\bibitem [{\citenamefont {Neill}\ \emph {et~al.}(2018)\citenamefont {Neill},
  \citenamefont {Roushan}, \citenamefont {Kechedzhi}, \citenamefont {Boixo},
  \citenamefont {Isakov}, \citenamefont {Smelyanskiy}, \citenamefont {Megrant},
  \citenamefont {Chiaro}, \citenamefont {Dunsworth}, \citenamefont {Arya},
  \citenamefont {Barends}, \citenamefont {Burkett}, \citenamefont {Chen},
  \citenamefont {Chen}, \citenamefont {Fowler}, \citenamefont {Foxen},
  \citenamefont {Giustina}, \citenamefont {Graff}, \citenamefont {Jeffrey},
  \citenamefont {Huang}, \citenamefont {Kelly}, \citenamefont {Klimov},
  \citenamefont {Lucero}, \citenamefont {Mutus}, \citenamefont {Neeley},
  \citenamefont {Quintana}, \citenamefont {Sank}, \citenamefont {Vainsencher},
  \citenamefont {Wenner}, \citenamefont {White}, \citenamefont {Neven},\ and\
  \citenamefont {Martinis}}]{Neill2018}%
  \BibitemOpen
  \bibfield  {author} {\bibinfo {author} {\bibfnamefont {C.}~\bibnamefont
  {Neill}}, \bibinfo {author} {\bibfnamefont {P.}~\bibnamefont {Roushan}},
  \bibinfo {author} {\bibfnamefont {K.}~\bibnamefont {Kechedzhi}}, \bibinfo
  {author} {\bibfnamefont {S.}~\bibnamefont {Boixo}}, \bibinfo {author}
  {\bibfnamefont {S.~V.}\ \bibnamefont {Isakov}}, \bibinfo {author}
  {\bibfnamefont {V.}~\bibnamefont {Smelyanskiy}}, \bibinfo {author}
  {\bibfnamefont {A.}~\bibnamefont {Megrant}}, \bibinfo {author} {\bibfnamefont
  {B.}~\bibnamefont {Chiaro}}, \bibinfo {author} {\bibfnamefont
  {A.}~\bibnamefont {Dunsworth}}, \bibinfo {author} {\bibfnamefont
  {K.}~\bibnamefont {Arya}}, \bibinfo {author} {\bibfnamefont {R.}~\bibnamefont
  {Barends}}, \bibinfo {author} {\bibfnamefont {B.}~\bibnamefont {Burkett}},
  \bibinfo {author} {\bibfnamefont {Y.}~\bibnamefont {Chen}}, \bibinfo {author}
  {\bibfnamefont {Z.}~\bibnamefont {Chen}}, \bibinfo {author} {\bibfnamefont
  {A.}~\bibnamefont {Fowler}}, \bibinfo {author} {\bibfnamefont
  {B.}~\bibnamefont {Foxen}}, \bibinfo {author} {\bibfnamefont
  {M.}~\bibnamefont {Giustina}}, \bibinfo {author} {\bibfnamefont
  {R.}~\bibnamefont {Graff}}, \bibinfo {author} {\bibfnamefont
  {E.}~\bibnamefont {Jeffrey}}, \bibinfo {author} {\bibfnamefont
  {T.}~\bibnamefont {Huang}}, \bibinfo {author} {\bibfnamefont
  {J.}~\bibnamefont {Kelly}}, \bibinfo {author} {\bibfnamefont
  {P.}~\bibnamefont {Klimov}}, \bibinfo {author} {\bibfnamefont
  {E.}~\bibnamefont {Lucero}}, \bibinfo {author} {\bibfnamefont
  {J.}~\bibnamefont {Mutus}}, \bibinfo {author} {\bibfnamefont
  {M.}~\bibnamefont {Neeley}}, \bibinfo {author} {\bibfnamefont
  {C.}~\bibnamefont {Quintana}}, \bibinfo {author} {\bibfnamefont
  {D.}~\bibnamefont {Sank}}, \bibinfo {author} {\bibfnamefont {A.}~\bibnamefont
  {Vainsencher}}, \bibinfo {author} {\bibfnamefont {J.}~\bibnamefont {Wenner}},
  \bibinfo {author} {\bibfnamefont {T.~C.}\ \bibnamefont {White}}, \bibinfo
  {author} {\bibfnamefont {H.}~\bibnamefont {Neven}},\ and\ \bibinfo {author}
  {\bibfnamefont {J.~M.}\ \bibnamefont {Martinis}},\ }\href
  {https://doi.org/10.1126/science.aao4309} {\bibfield  {journal} {\bibinfo
  {journal} {Science}\ }\textbf {\bibinfo {volume} {360}},\ \bibinfo {pages}
  {195} (\bibinfo {year} {2018})}\BibitemShut {NoStop}%
\bibitem [{\citenamefont {Arute}\ \emph {et~al.}(2019)\citenamefont {Arute},
  \citenamefont {Arya}, \citenamefont {Babbush}, \citenamefont {Bacon},
  \citenamefont {Bardin}, \citenamefont {Barends}, \citenamefont {Biswas},
  \citenamefont {Boixo}, \citenamefont {Brandao}, \citenamefont {Buell},
  \citenamefont {Burkett}, \citenamefont {Chen}, \citenamefont {Chen},
  \citenamefont {Chiaro}, \citenamefont {Collins}, \citenamefont {Courtney},
  \citenamefont {Dunsworth}, \citenamefont {Farhi}, \citenamefont {Foxen},
  \citenamefont {Fowler}, \citenamefont {Gidney}, \citenamefont {Giustina},
  \citenamefont {Graff}, \citenamefont {Guerin}, \citenamefont {Habegger},
  \citenamefont {Harrigan}, \citenamefont {Hartmann}, \citenamefont {Ho},
  \citenamefont {Hoffmann}, \citenamefont {Huang}, \citenamefont {Humble},
  \citenamefont {Isakov}, \citenamefont {Jeffrey}, \citenamefont {Jiang},
  \citenamefont {Kafri}, \citenamefont {Kechedzhi}, \citenamefont {Kelly},
  \citenamefont {Klimov}, \citenamefont {Knysh}, \citenamefont {Korotkov},
  \citenamefont {Kostritsa}, \citenamefont {Landhuis}, \citenamefont
  {Lindmark}, \citenamefont {Lucero}, \citenamefont {Lyakh}, \citenamefont
  {Mandr{\`{a}}}, \citenamefont {McClean}, \citenamefont {McEwen},
  \citenamefont {Megrant}, \citenamefont {Mi}, \citenamefont {Michielsen},
  \citenamefont {Mohseni}, \citenamefont {Mutus}, \citenamefont {Naaman},
  \citenamefont {Neeley}, \citenamefont {Neill}, \citenamefont {Niu},
  \citenamefont {Ostby}, \citenamefont {Petukhov}, \citenamefont {Platt},
  \citenamefont {Quintana}, \citenamefont {Rieffel}, \citenamefont {Roushan},
  \citenamefont {Rubin}, \citenamefont {Sank}, \citenamefont {Satzinger},
  \citenamefont {Smelyanskiy}, \citenamefont {Sung}, \citenamefont
  {Trevithick}, \citenamefont {Vainsencher}, \citenamefont {Villalonga},
  \citenamefont {White}, \citenamefont {Yao}, \citenamefont {Yeh},
  \citenamefont {Zalcman}, \citenamefont {Neven},\ and\ \citenamefont
  {Martinis}}]{Arute2019}%
  \BibitemOpen
  \bibfield  {author} {\bibinfo {author} {\bibfnamefont {F.}~\bibnamefont
  {Arute}}, \bibinfo {author} {\bibfnamefont {K.}~\bibnamefont {Arya}},
  \bibinfo {author} {\bibfnamefont {R.}~\bibnamefont {Babbush}}, \bibinfo
  {author} {\bibfnamefont {D.}~\bibnamefont {Bacon}}, \bibinfo {author}
  {\bibfnamefont {J.~C.}\ \bibnamefont {Bardin}}, \bibinfo {author}
  {\bibfnamefont {R.}~\bibnamefont {Barends}}, \bibinfo {author} {\bibfnamefont
  {R.}~\bibnamefont {Biswas}}, \bibinfo {author} {\bibfnamefont
  {S.}~\bibnamefont {Boixo}}, \bibinfo {author} {\bibfnamefont {F.~G. S.~L.}\
  \bibnamefont {Brandao}}, \bibinfo {author} {\bibfnamefont {D.~A.}\
  \bibnamefont {Buell}}, \bibinfo {author} {\bibfnamefont {B.}~\bibnamefont
  {Burkett}}, \bibinfo {author} {\bibfnamefont {Y.}~\bibnamefont {Chen}},
  \bibinfo {author} {\bibfnamefont {Z.}~\bibnamefont {Chen}}, \bibinfo {author}
  {\bibfnamefont {B.}~\bibnamefont {Chiaro}}, \bibinfo {author} {\bibfnamefont
  {R.}~\bibnamefont {Collins}}, \bibinfo {author} {\bibfnamefont
  {W.}~\bibnamefont {Courtney}}, \bibinfo {author} {\bibfnamefont
  {A.}~\bibnamefont {Dunsworth}}, \bibinfo {author} {\bibfnamefont
  {E.}~\bibnamefont {Farhi}}, \bibinfo {author} {\bibfnamefont
  {B.}~\bibnamefont {Foxen}}, \bibinfo {author} {\bibfnamefont
  {A.}~\bibnamefont {Fowler}}, \bibinfo {author} {\bibfnamefont
  {C.}~\bibnamefont {Gidney}}, \bibinfo {author} {\bibfnamefont
  {M.}~\bibnamefont {Giustina}}, \bibinfo {author} {\bibfnamefont
  {R.}~\bibnamefont {Graff}}, \bibinfo {author} {\bibfnamefont
  {K.}~\bibnamefont {Guerin}}, \bibinfo {author} {\bibfnamefont
  {S.}~\bibnamefont {Habegger}}, \bibinfo {author} {\bibfnamefont {M.~P.}\
  \bibnamefont {Harrigan}}, \bibinfo {author} {\bibfnamefont {M.~J.}\
  \bibnamefont {Hartmann}}, \bibinfo {author} {\bibfnamefont {A.}~\bibnamefont
  {Ho}}, \bibinfo {author} {\bibfnamefont {M.}~\bibnamefont {Hoffmann}},
  \bibinfo {author} {\bibfnamefont {T.}~\bibnamefont {Huang}}, \bibinfo
  {author} {\bibfnamefont {T.~S.}\ \bibnamefont {Humble}}, \bibinfo {author}
  {\bibfnamefont {S.~V.}\ \bibnamefont {Isakov}}, \bibinfo {author}
  {\bibfnamefont {E.}~\bibnamefont {Jeffrey}}, \bibinfo {author} {\bibfnamefont
  {Z.}~\bibnamefont {Jiang}}, \bibinfo {author} {\bibfnamefont
  {D.}~\bibnamefont {Kafri}}, \bibinfo {author} {\bibfnamefont
  {K.}~\bibnamefont {Kechedzhi}}, \bibinfo {author} {\bibfnamefont
  {J.}~\bibnamefont {Kelly}}, \bibinfo {author} {\bibfnamefont {P.~V.}\
  \bibnamefont {Klimov}}, \bibinfo {author} {\bibfnamefont {S.}~\bibnamefont
  {Knysh}}, \bibinfo {author} {\bibfnamefont {A.}~\bibnamefont {Korotkov}},
  \bibinfo {author} {\bibfnamefont {F.}~\bibnamefont {Kostritsa}}, \bibinfo
  {author} {\bibfnamefont {D.}~\bibnamefont {Landhuis}}, \bibinfo {author}
  {\bibfnamefont {M.}~\bibnamefont {Lindmark}}, \bibinfo {author}
  {\bibfnamefont {E.}~\bibnamefont {Lucero}}, \bibinfo {author} {\bibfnamefont
  {D.}~\bibnamefont {Lyakh}}, \bibinfo {author} {\bibfnamefont
  {S.}~\bibnamefont {Mandr{\`{a}}}}, \bibinfo {author} {\bibfnamefont {J.~R.}\
  \bibnamefont {McClean}}, \bibinfo {author} {\bibfnamefont {M.}~\bibnamefont
  {McEwen}}, \bibinfo {author} {\bibfnamefont {A.}~\bibnamefont {Megrant}},
  \bibinfo {author} {\bibfnamefont {X.}~\bibnamefont {Mi}}, \bibinfo {author}
  {\bibfnamefont {K.}~\bibnamefont {Michielsen}}, \bibinfo {author}
  {\bibfnamefont {M.}~\bibnamefont {Mohseni}}, \bibinfo {author} {\bibfnamefont
  {J.}~\bibnamefont {Mutus}}, \bibinfo {author} {\bibfnamefont
  {O.}~\bibnamefont {Naaman}}, \bibinfo {author} {\bibfnamefont
  {M.}~\bibnamefont {Neeley}}, \bibinfo {author} {\bibfnamefont
  {C.}~\bibnamefont {Neill}}, \bibinfo {author} {\bibfnamefont {M.~Y.}\
  \bibnamefont {Niu}}, \bibinfo {author} {\bibfnamefont {E.}~\bibnamefont
  {Ostby}}, \bibinfo {author} {\bibfnamefont {A.}~\bibnamefont {Petukhov}},
  \bibinfo {author} {\bibfnamefont {J.~C.}\ \bibnamefont {Platt}}, \bibinfo
  {author} {\bibfnamefont {C.}~\bibnamefont {Quintana}}, \bibinfo {author}
  {\bibfnamefont {E.~G.}\ \bibnamefont {Rieffel}}, \bibinfo {author}
  {\bibfnamefont {P.}~\bibnamefont {Roushan}}, \bibinfo {author} {\bibfnamefont
  {N.~C.}\ \bibnamefont {Rubin}}, \bibinfo {author} {\bibfnamefont
  {D.}~\bibnamefont {Sank}}, \bibinfo {author} {\bibfnamefont {K.~J.}\
  \bibnamefont {Satzinger}}, \bibinfo {author} {\bibfnamefont {V.}~\bibnamefont
  {Smelyanskiy}}, \bibinfo {author} {\bibfnamefont {K.~J.}\ \bibnamefont
  {Sung}}, \bibinfo {author} {\bibfnamefont {M.~D.}\ \bibnamefont
  {Trevithick}}, \bibinfo {author} {\bibfnamefont {A.}~\bibnamefont
  {Vainsencher}}, \bibinfo {author} {\bibfnamefont {B.}~\bibnamefont
  {Villalonga}}, \bibinfo {author} {\bibfnamefont {T.}~\bibnamefont {White}},
  \bibinfo {author} {\bibfnamefont {Z.~J.}\ \bibnamefont {Yao}}, \bibinfo
  {author} {\bibfnamefont {P.}~\bibnamefont {Yeh}}, \bibinfo {author}
  {\bibfnamefont {A.}~\bibnamefont {Zalcman}}, \bibinfo {author} {\bibfnamefont
  {H.}~\bibnamefont {Neven}},\ and\ \bibinfo {author} {\bibfnamefont {J.~M.}\
  \bibnamefont {Martinis}},\ }\href {https://doi.org/10.1038/s41586-019-1666-5}
  {\bibfield  {journal} {\bibinfo  {journal} {Nature}\ }\textbf {\bibinfo
  {volume} {574}},\ \bibinfo {pages} {505} (\bibinfo {year}
  {2019})}\BibitemShut {NoStop}%
\bibitem [{\citenamefont {O'Malley}\ \emph {et~al.}(2015)\citenamefont
  {O'Malley}, \citenamefont {Kelly}, \citenamefont {Barends}, \citenamefont
  {Campbell}, \citenamefont {Chen}, \citenamefont {Chen}, \citenamefont
  {Chiaro}, \citenamefont {Dunsworth}, \citenamefont {Fowler}, \citenamefont
  {Hoi}, \citenamefont {Jeffrey}, \citenamefont {Megrant}, \citenamefont
  {Mutus}, \citenamefont {Neill}, \citenamefont {Quintana}, \citenamefont
  {Roushan}, \citenamefont {Sank}, \citenamefont {Vainsencher}, \citenamefont
  {Wenner}, \citenamefont {White}, \citenamefont {Korotkov}, \citenamefont
  {Cleland},\ and\ \citenamefont {Martinis}}]{OMalley2015}%
  \BibitemOpen
  \bibfield  {author} {\bibinfo {author} {\bibfnamefont {P.~J.~J.}\
  \bibnamefont {O'Malley}}, \bibinfo {author} {\bibfnamefont {J.}~\bibnamefont
  {Kelly}}, \bibinfo {author} {\bibfnamefont {R.}~\bibnamefont {Barends}},
  \bibinfo {author} {\bibfnamefont {B.}~\bibnamefont {Campbell}}, \bibinfo
  {author} {\bibfnamefont {Y.}~\bibnamefont {Chen}}, \bibinfo {author}
  {\bibfnamefont {Z.}~\bibnamefont {Chen}}, \bibinfo {author} {\bibfnamefont
  {B.}~\bibnamefont {Chiaro}}, \bibinfo {author} {\bibfnamefont
  {A.}~\bibnamefont {Dunsworth}}, \bibinfo {author} {\bibfnamefont {A.~G.}\
  \bibnamefont {Fowler}}, \bibinfo {author} {\bibfnamefont {I.-C.}\
  \bibnamefont {Hoi}}, \bibinfo {author} {\bibfnamefont {E.}~\bibnamefont
  {Jeffrey}}, \bibinfo {author} {\bibfnamefont {A.}~\bibnamefont {Megrant}},
  \bibinfo {author} {\bibfnamefont {J.}~\bibnamefont {Mutus}}, \bibinfo
  {author} {\bibfnamefont {C.}~\bibnamefont {Neill}}, \bibinfo {author}
  {\bibfnamefont {C.}~\bibnamefont {Quintana}}, \bibinfo {author}
  {\bibfnamefont {P.}~\bibnamefont {Roushan}}, \bibinfo {author} {\bibfnamefont
  {D.}~\bibnamefont {Sank}}, \bibinfo {author} {\bibfnamefont {A.}~\bibnamefont
  {Vainsencher}}, \bibinfo {author} {\bibfnamefont {J.}~\bibnamefont {Wenner}},
  \bibinfo {author} {\bibfnamefont {T.~C.}\ \bibnamefont {White}}, \bibinfo
  {author} {\bibfnamefont {A.~N.}\ \bibnamefont {Korotkov}}, \bibinfo {author}
  {\bibfnamefont {A.~N.}\ \bibnamefont {Cleland}},\ and\ \bibinfo {author}
  {\bibfnamefont {J.~M.}\ \bibnamefont {Martinis}},\ }\href
  {https://doi.org/10.1103/PhysRevApplied.3.044009} {\bibfield  {journal}
  {\bibinfo  {journal} {Phys. Rev. Appl.}\ }\textbf {\bibinfo {volume} {3}},\
  \bibinfo {pages} {044009} (\bibinfo {year} {2015})}\BibitemShut {NoStop}%
\end{thebibliography}%


%
\end{document}


\title{Supplementary Material for \\
Native approach to controlled-Z gates in inductively coupled fluxonium qubits}

\author{Xizheng Ma}
\thanks{These authors contributed equally to this work}
\affiliation{DAMO Quantum Laboratory, Alibaba Group, Hangzhou, Zhejiang 311121, China}
\author{Gengyan Zhang}
\thanks{These authors contributed equally to this work}
\author{Feng Wu}
\thanks{These authors contributed equally to this work}
\author{Feng Bao}
\author{Xu Chang}
\author{Jianjun Chen}
\thanks{Current address: Xinxiao Electronics Inc., Hangzhou, China}
\author{Hao Deng}
\author{Ran Gao}
\affiliation{DAMO Quantum Laboratory, Alibaba Group, Hangzhou, Zhejiang 311121, China}
\author{Xun Gao}
\affiliation{DAMO Quantum Laboratory, Alibaba Group USA, Bellevue, WA 98004, USA}
\author{Lijuan Hu}
\affiliation{DAMO Quantum Laboratory, Alibaba Group, Hangzhou, Zhejiang 311121, China}
\author{Honghong Ji}
\author{Hsiang-Sheng Ku}
\thanks{Current address: IQM Quantum Computers, 80992 Munich, Germany}
\author{Kannan Lu}
\author{Lu Ma}
\author{Liyong Mao}
\author{Zhijun Song}
\author{Hantao Sun}
\author{Chengchun Tang}
\author{Fei Wang}
\author{Hongcheng Wang}
\author{Tenghui Wang}
\author{Tian Xia}
\author{Make Ying}
\author{Huijuan Zhan}
\author{Tao Zhou}
\author{Mengyu Zhu}
\author{Qingbin Zhu}
\affiliation{DAMO Quantum Laboratory, Alibaba Group, Hangzhou, Zhejiang 311121, China}
\author{Yaoyun Shi}
\affiliation{DAMO Quantum Laboratory, Alibaba Group USA, Bellevue, WA 98004, USA}
\author{Hui-Hai Zhao}
\email{huihai.zhh@alibaba-inc.com}
\affiliation{DAMO Quantum Laboratory, Alibaba Group, Beijing 100102, China}
\author{Chunqing Deng}
\email{chunqing.cd@alibaba-inc.com}
\affiliation{DAMO Quantum Laboratory, Alibaba Group, Hangzhou, Zhejiang 311121, China}

\maketitle

\bookmarksetup{startatroot}

\section{Full Hamiltonian and coupling strength}\label{sec:general}

The coupled system investigated in this work is accurately described by the full Hamiltonian,
\begin{equation}\label{eqn:full-Hamiltonian}
\begin{aligned}
    H &= \sum_{i = A,B}H_0^i(\Phi_i) + H_I(\Phi_A,\Phi_B)\\
    &=\sum_{i = A,B} \bigg\{ E_c^i \hat{n}_i^2 + E_J^i \cos{\hat{\varphi}_i} + E_L^i\left(\hat{\varphi}_i - 2\pi \frac{\Phi_i}{\Phi_0}\right)^2 \bigg\} + \mathcal{J}\left(\hat{\varphi}_A - 2\pi \frac{\Phi_A}{\Phi_0}\right)\left(\hat{\varphi}_B - 2\pi \frac{\Phi_B}{\Phi_0}\right),
\end{aligned}
\end{equation}
where $E_c^i$, $E_L^i$, $E_J^i$ are respectively the charging, inductive, and Josephson energy of qubit $i$, described by its Cooper-pair number operator $\hat{n}_i$ and the reduced superconducting phase operator $\hat{\varphi}_i$. The phase operator $\hat{\varphi}_i$ relates to the current operator $\hat{I}_i$ of the main text through the loop inductance $L_i$ of the qubit according to $\hat{\varphi}_i \Phi_0/2\pi = \hat{I}_i L_i$. Written in the phase basis, the inductive coupling strength is $\mathcal{J} = M(\Phi_0/2\pi)^2/L_AL_B$, independent of the external flux observed by the qubits. 
\autoref{tab:device_params} summarizes the measured values of these parameters. 

\begin{table}[h]
    \begin{tabular}{|M{4em} || M{5em}|M{5em}|M{5em}|M{5em}|}
    \hline 
    \multirow{2}{*}{ Qubit } & $E_{C}/h$ & $E_{L}/h$ & $E_{J}/h$ & $\mathcal{J}/h$ \\
    &(GHz) & (GHz) & (GHz) & (MHz)\\
    \hline
    $A$ & $1.61$ & $0.45$ & $2.89$ & \multirow{2}{*}{$3.5$} \\
    $B$ & $1.24$ & $0.45$ & $2.68$ & \\
    \hline 
    \end{tabular}
    \caption{Measured parameters of the coupled system.}
    \label{tab:device_params}
\end{table}

\begin{figure}[h]
    \centering
    \includegraphics{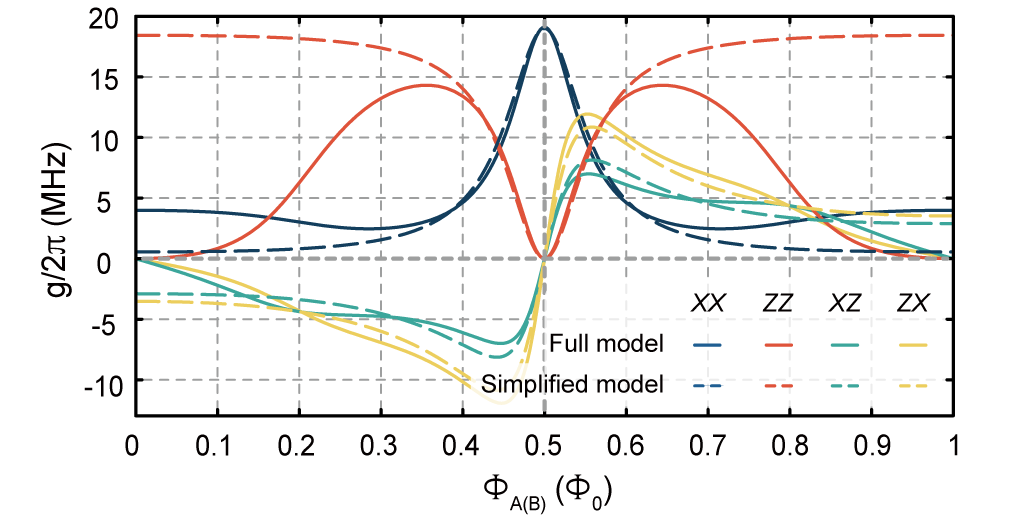}
    \caption{Taking $\Phi_A = \Phi_B = \Phi$ and using the measured parameters of \autoref{tab:device_params}, we plot the coupling strengths $g_{xx}$, $g_{zz}$, $g_{xz}$ and $g_{zx}$. Solid lines are the coupling strength predicted by the full model (\autoref{eqn:coupling_strengths}), and dashed lines are the coupling strength predicted by the simplified model given in Eq.~(2) of the main text.}
    \label{fig:coupling_landscape}
\end{figure}

Defining the lowest two energy eigenstates of $H_0^i(\Phi_i)$ as qubit $i$ at external flux $\Phi_i$, we can rewrite the interaction Hamiltonian of \autoref{eqn:full-Hamiltonian} into a general form,
\begin{equation}\label{eqn:coupling_Hamiltonian}
    H_I(\Phi_A,\Phi_B)/\hbar = g_{xx} \hatsigma{x}^A \hatsigma{x}^B + g_{zz} \hatsigma{z}^A \hatsigma{z}^B + g_{xz} \hatsigma{x}^A \hatsigma{z}^B + g_{zx} \hatsigma{z}^A \hatsigma{x}^B,
\end{equation}
where $\hatsigma{x}^i$ and $\hatsigma{z}^i$ are the pauli operators of qubit $i$, and
\begin{equation}\label{eqn:coupling_strengths}
    \begin{aligned}
        g_{xx} &= \frac{\mathcal{J}}{4\hbar}\bigg( \bra{g}\hat{\varphi}_A\ket{e} + \bra{e}\hat{\varphi}_A\ket{g}\bigg)\bigg( \bra{g}\hat{\varphi}_B\ket{e} + \bra{e}\hat{\varphi}_B\ket{g}\bigg)\\
        g_{zz} &= \frac{\mathcal{J}}{4\hbar}\bigg( \bra{e}\hat{\varphi}_A\ket{e} - \bra{g}\hat{\varphi}_A\ket{g}\bigg)\bigg( \bra{e}\hat{\varphi}_B\ket{e} - \bra{g}\hat{\varphi}_B\ket{g}\bigg)\\
        g_{xz} &= \frac{\mathcal{J}}{4\hbar}\bigg( \bra{g}\hat{\varphi}_A\ket{e} + \bra{e}\hat{\varphi}_A\ket{g}\bigg)\bigg( \bra{e}\hat{\varphi}_B\ket{e} - \bra{g}\hat{\varphi}_B\ket{g}\bigg)\\
        g_{zx} &= \frac{\mathcal{J}}{4\hbar}\bigg( \bra{e}\hat{\varphi}_A\ket{e} - \bra{g}\hat{\varphi}_A\ket{g}\bigg)\bigg( \bra{g}\hat{\varphi}_B\ket{e} + \bra{e}\hat{\varphi}_B\ket{g}\bigg)
    \end{aligned}
\end{equation}
are the interaction strengths. Using the parameters of \autoref{tab:device_params} and taking $\Phi_A = \Phi_B = \Phi$, \autoref{fig:coupling_landscape} illustrates the dependence of these coupling strengths as a function of the applied external flux. In comparison, we also plot in dashed lines the coupling strengths predicted by the simplified model given in Eq.~(2) of the main text. Although the simplified model is only accurate when $\Phi_A$ and $\Phi_B$ are both close to $\Phi_0/2$, the physical intuition it provides remains true: by moving both qubits away from their degeneracy position, a native \textit{ZZ}-interaction can be adiabatically switched on.

\section{the idle position}

\begin{figure}[!]
    \centering
    \includegraphics{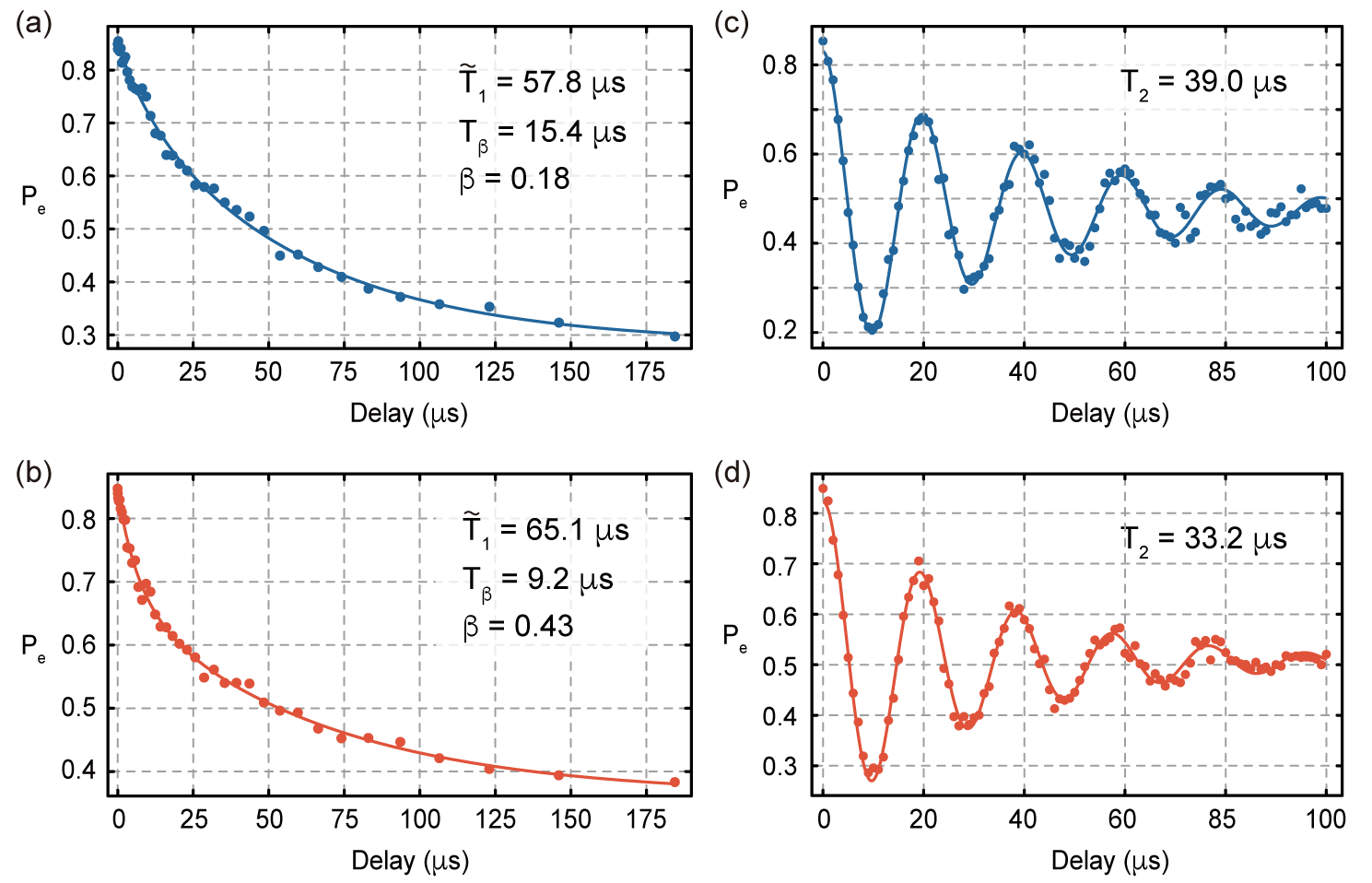}
    \caption{Characteristic energy decay functions are shown for both qubit A (a) and qubit B (b). These functions frequently display features of double-exponential decay, which we account for by fitting the data according to $\braket{P_e(t)} \propto e^{\beta \big(\text{exp}(-t/T_\beta)-1\big)}e^{-t/\tilde{T}_1}$~\cite{pop2014,Gustavsson2016}. We also measure the Ramsey decoherence times of both qubit A (c) and qubit B (d).}
    \label{fig:coherence_SP}
\end{figure}

The idle position, with both qubits parked at $\Phi_0/2$, is of particular interest to the operation of the coupled system.
With all native interactions except $g_{xx}$ switched off, this idle position is ideal for performing qubit readout~\cite{Bao2022} and single qubit operations. \autoref{fig:coherence_SP} shows the coherence times measured at the idle position. Compared to previous demonstrations of excellent coherence times~\cite{pop2014,Zhang2021,Somoroff2023}, the performance of our qubits at the flux degeneracy positions are rather 
underwhelming. In particular, the $T_1$ measurements show symptoms typical of systems afflicted by quasiparticle~\cite{pop2014,Gustavsson2016} or TLS~\cite{Barends2013,Klimov2018} poisoning, and are not only subjected to abrupt fluctuations in time, but also frequently display double-exponential decay. Such time-fluctuations of the qubit decoherence posses a challenge for estimating the decoherence limits of our gate operations, which we address in Sec.~\ref{sec:SQ-gate} and Sec.~\ref{sec:gate_error}.

\section{Qubit decoherence from undesirable harmonic modes}

\begin{figure}[!]
    \centering
    \includegraphics{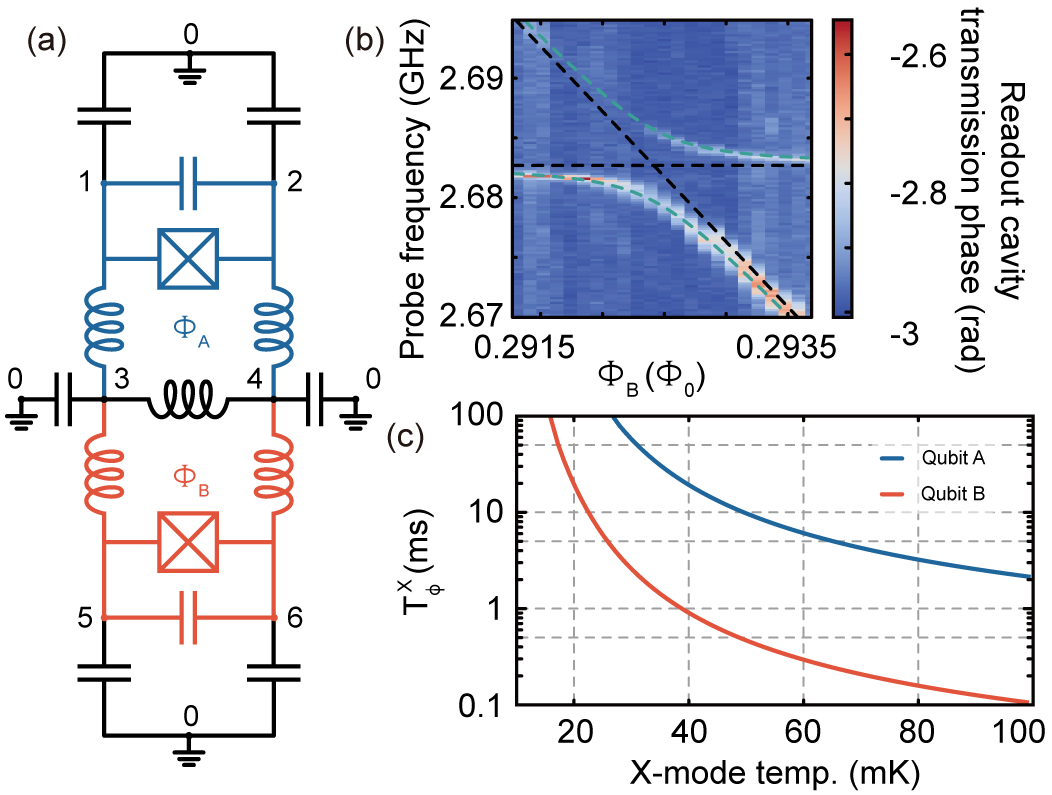}
    \caption{
    (a) The galvanically connected fluxonium pair is described by the full circuit model. All circuit nodes are numbered, and the two qubit modes A and B are colored blue and red respectively. 
    (b) When qubit B comes into resonance with the X-mode located at $2.683~\text{GHz}$, an avoided crossing approximately $2\pi\times 6~\text{MHz}$ wide can be observed. Based on this measurement, we calculate a dispersive shift $2\chi_{B}\approx 2\pi\times 64~\text{kHz}$ per X-mode photon on qubit B when it is kept at $\Phi_0/2$
    (c) Using the extracted dispersive shifts, we plot the X-mode induced additional dephasing $T_\phi^{X}$ as a function of X-mode temperature for both qubit A and B, assuming a X-mode loss rate $\kappa \approx 2\pi\times 500~\text{kHz}$.}
    \label{fig:xmode_diagram}
\end{figure}

Because our floating fluxnoium qubits are connected galvanically, our system contains a series of undesirable harmonic modes that could cause additional qubit decoherence. These modes and their coupling to the qubits can be found exactly by analyzing the full circuit model (\autoref{fig:xmode_diagram}(a)) according to Ref.~\cite{Ding2021}. Here, we provide a brief description on the lowest frequency mode (X-mode), which is most detrimental to the qubit coherence times.

The X-mode is a harmonic mode where currents oscillate up-and-down in the circuit of \autoref{fig:xmode_diagram}(a), and can be easily understood by performing $\Delta$-Y transformations on the capacitive sub-circuits confined by nodes $(0,1,2)$ and nodes $(0,5,6)$.
Through their shared capacitive networks, this mode could couple strongly to the qubits, where qubit A corresponds to horizontal oscillations between nodes $(1,2)$ and qubit B that between nodes $(5,6)$.
For our qubit parameters, the X-mode typically resides around $2.7~\text{GHz}$, a frequency low enough such that its thermal fluctuations could induce significant qubit decoherence. One way to combat this problem is to eliminate the X-mode altogether either by breaking the galvanic connection using a geometric inductance or by moving toward a grounded circuit design.

In this work, we pursue the alternative approach where we suppress the couplings between the X-mode and the qubits through a carefully designed circuit symmetry. Specifically, we ensure the oscillating current associated with the X-mode induces approximately equal electrostatic potential on nodes $5$ and $6$, and thereby reduces the coupling between qubit B and X-mode. Indeed, \autoref{fig:xmode_diagram}(b) shows the avoided crossing when qubit B comes into resonance with the X-mode at $2.683~\text{GHz}$, from which we extract a small dispersive shift $2\chi_{B}\approx 2\pi\times 64~\text{kHz}$ per X-mode photon on qubit B when it is kept at $\Phi_0/2$. Doing the same for nodes $1$ and $2$, we similarly suppress the coupling between qubit A and the X-mode, which we bound to $2\chi_A \lesssim 2\pi \times 7~\text{kHz}$. Assuming an X-mode loss rate $\kappa \approx 2\pi\times 500~\text{kHz}$ based on our typical junction array loss rate, \autoref{fig:xmode_diagram}(c) plots the X-mode induced additional dephasing $T_\phi^{X}$ as a function of X-mode temperature~\cite{Zhang2017}. For reasonable temperatures below $100~\text{mK}$, we find $T_\phi^{X}$ to be largely negligible. 

\section{Effects of the native \textit{XZ}-interaction}\label{sec:XZ}

Analogous to cross-resonance schemes~\cite{Rigetti2010}, the native \textit{XZ}-interaction enables fast controlled-NOT gates when the coupling strength is modulated~\cite{Didier2015} at the appropriate frequency. 
To see this, we examine a simplified model with a time-dependent Hamiltonian in the rotating frame of qubit B,
\begin{equation} \label{eqn:XZ-interaction}
    H^\prime/\hbar = \frac{\omega_A}{2} \hatsigma{z}^A + g_{xz}(t) \hatsigma{x}^A \hatsigma{z}^B.
\end{equation}
As we explained in the main text and Sec.~\ref{sec:general}, we can control $g_{xz}$ \textit{in-situ} using the external flux $\Phi_i$ applied to the qubit loops.
When this coupling strength is sinusoidally modulated at the frequency of qubit A, $g_{xz}(t) = \tilde{g}_{xz}\sin{(\omega_B t)}$, \autoref{eqn:XZ-interaction} describes qubit A being resonantly driven with a Rabi rate $\Omega_R = \tilde{g}_{xz}\hatsigma{z}^B$ that is dependent on the state of qubit B. Consequently, a controlled-NOT gate is realized when this interaction is turned on for a duration $t = 2\pi/\tilde{g}_{xz}$, inverting qubit A's population while picking up a sign that is dependent on the state of qubit B.

In this work, however, such single-qubit rotations appear as control errors to our two-qubit CZ gates. While the precise amount of errors can be found explicitly by solving the Rabi model in \autoref{eqn:XZ-interaction}, doing so reveals little insight. Instead, because we modulate the external flux at a frequency that is much smaller than either qubit's resonant frequency ($\omega_m \ll \omega_{A,B}$), the problem of finding the \textit{XZ}-error can be recast~\cite{Amin2009} into how much our sinusoidal modulations violate the adiabatic limit of the coupled system. Indeed, the adiabatic theorem postulates that our coupled two-qubit system shall remain in the same instantaneous energy eigenstate under adequately slow modulations of $\Phiext{A,B}(t)$, and therefore accumulate no population error when $\Phiext{A,B}(t)$ are returned to their initial values at the end of the modulation. For this reason, it is important that we modulate our external flux using $\sin{(\omega_m t)}$ instead of $\cos{(\omega_m t)}$ such that we always start and return the qubits adiabatically to their respective flux-degeneracy positions.
As we detail in Sec.~\ref{sec:LandauZener}, we ensure a negligible \textit{XZ}-error below $10^{-4}$ on either qubit by choosing a small $\omega_m = 2\pi\times 50~\text{MHz}$ and keeping the flux modulation amplitude within $\pm 0.15~\Phi_0$.

\section{Effects of the \textit{XX}-interaction}\label{sec:XX}

In the context of our adiabatic CZ gate, the native \textit{XX}-interaction could lead to control errors in two ways.

First, when the $XX$-interaction strength is modulated at a frequency close to the sum or difference frequency of the two qubits, a parametric interaction~\cite{Bertet2006, Niskanen2006, Niskanen2007, Weiss2022} could lead to undesirable qubit rotations.
Similar to the \textit{XZ}-interactions, by ensuring $\omega_m\ll \omega_A(t) \pm \omega_B(t)$ for all time, we suppress this error by modulating the coupled system within its adiabatic limit (see Sec.~\ref{sec:LandauZener}).

\begin{figure}[h]
    \centering
    \includegraphics{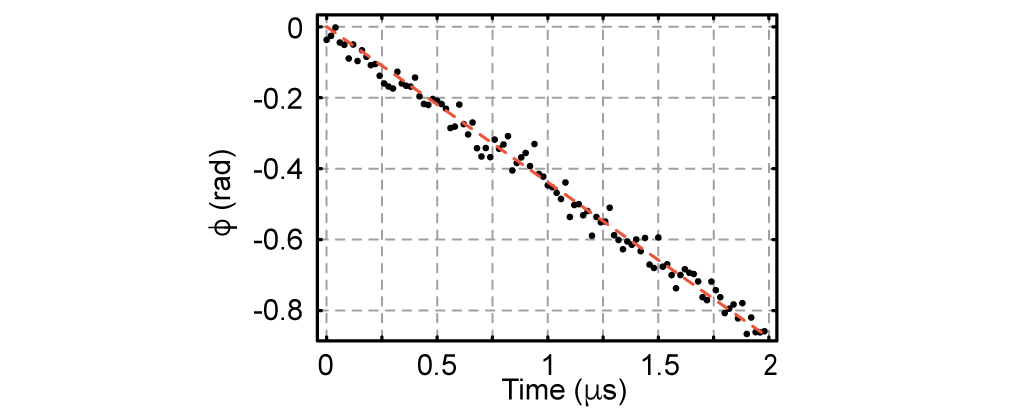}
    \caption{The conditional phase ($\phi$) accumulated in qubit A is measured as a function of time. A linear fit (dashed line) through the origin find the accumulation speed $v_\phi \approx -0.438 \text{rad}/\mu\text{s}$ }
    \label{fig:residalZZ}
\end{figure}

In addition to the non-adiabatic errors, the $XX$-interaction between the computational states and higher-energy qubit states also results in a small residual \textit{ZZ}-interaction at the idle position~\cite{Ficheux2021,Nguyen2022}. Although it does not cause errors to the CZ gate, this residual \textit{ZZ}-interaction causes uncertainties in the qubit frequencies that incur phase errors during our single-qubit operations. Here, we explicitly measure this residual \textit{ZZ}-interaction strength using the accumulation speed of conditional phase on qubit A. Specifically, we keep both qubits at $\Phi_0/2$ and use the Ramsey-type experiment described in the main text to measure the difference in the quantum phases accumulated in qubit A when qubit B is respectively prepared in $\ket{g}$ and $\ket{e}$ state. As shown in \autoref{fig:residalZZ}, this conditional phase ($\phi$) accumulates linearly at a constant speed $v_\phi \approx -0.438 \text{rad}/\mu\text{s}$ and is directly related to $g_{zz}^\text{res}$ according to,
\begin{equation}
    g_{zz}^\text{res} = \frac{v_\phi}{4} \approx 2\pi \times 17.4~\text{kHz}.
\end{equation}
For our $15~\text{ns}$ single-qubit operations performed with both qubits placed at $\Phi_0/2$, we therefore estimate $g_{zz}^\text{res}$ to contribute a negligible error below $10^{-5}$. 

\section{Single qubit gate}\label{sec:SQ-gate}

We perform single-qubit operations at the idle position with both qubits parked at $\Phi_0/2$. Using the procedures discussed in Ref.~\cite{Bao2022} and a fixed gate duration of $15~\text{ns}$, we calibrate for each qubit a set of primary gate operations $\{I,X_\pi,Y_\pi,X_{\pm \pi/2},Y_{\pm \pi/2}\}$, consisting of single-qubit rotations of angle $0$, $\pi$, and $\pi/2$ around two independent axes $X$ and $Y$. 

\begin{figure}
    \centering
    \includegraphics{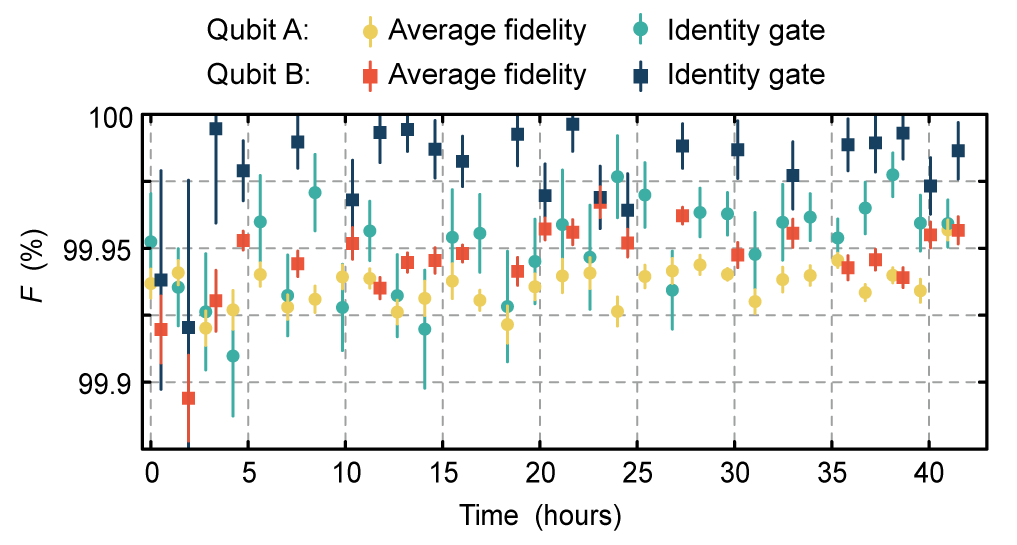}
    \caption{Using standard RB, we monitor the fidelity of the single-qubit gates over a period of 40 hours. The error bars correspond to fitting uncertainties.}
    \label{fig:sq_fidelity}
\end{figure}

We characterize the quality of these single-qubit gates using standard RB techniques~\cite{Magesan2012,Barends2014}. Monitored over $40$ hours (\autoref{fig:sq_fidelity}), we find an average fidelity $F_A = 99.94(1) \pm 0.00(5) \%$ across all primary gates on qubit A and $F_B = 99.94(7) \pm 0.00(7) \%$ on qubit B, where the uncertainty intervals capture the time fluctuations in the measured fidelities. Because our qubit system suffers from fluctuating coherence times, likely caused by nearby TLS with fluctuating frequencies, we refrain from estimating the decoherence limit of these gates based on the measured $T_1$ and $T_2$ values. Instead, we explicitly characterize the decoherence limit using the fidelity of the identity gates $I$ to find $F_A^{I} = 99.95(0) \pm 0.01(7) \%$ and $F_B^{I} = 99.97(9) \pm 0.01(8)\%$. The large uncertainties on these decoherence limits corroborate the fluctuating coherence times observed by the qubits.

\section{Dynamical decoupling}

Our CZ gate scheme requires us to bring both qubits away from the flux degeneracy position, which leads to increased sensitivity to flux noise and reduced coherence times.
To combat this, we perform a continuous version of dynamical decoupling, where we sinusoidally modulate the external flux $\Phiext{}$ applied to the qubit around $\Phi_0/2$. In Sec.~\ref{sec:decoupling_theory}, we justify our dynamical decoupling scheme for an environment dominated by $1/f$-noise by deriving its noise-frequency-selection characteristics. We then demonstrate the  efficacy of our scheme by measuring the qubit coherence times in Sec.~\ref{sec:decoupling_experiment}.

\subsection{Theory}\label{sec:decoupling_theory}

We show how a classical flux noise $\Phi_n(t)$ leads to qubit dephasing by modelling it as a sum of sinusoidal signals at equally-spaced frequencies $\omega_i$ with random phase $\xi_i$ uniformly distributed within range $[0,2\pi]$,
\begin{equation}
    \Phi_n(t) = \sum_i \Phi_i(\omega_i) \sin{(\omega_i t + \xi_i)},
\end{equation}
where the noise amplitude $\Phi_i(\omega_i)$ is related to the double-sided noise-power spectral density $S_{\Phi}(\omega)$ 
through an integration over frequency range $2\delta \omega = \omega_{i+1} - \omega_{i}$,
\begin{equation}
    \Phi_i^2(\omega_i) = \frac{2}{\pi} \int_{\omega_i-\delta\omega}^{\omega_i+\delta\omega} d\omega S_{\Phi}(\omega).
\end{equation}
When the qubit is subjected to a time-dependent flux control $\Phi(t)$, it observes an overall external flux $\Phiext{\text{tot}}(t) = \Phi(t) + \Phi_n(t)$ that leads to a noisy time-dependent qubit frequency. Treating the flux noise as a small perturbation, this noisy qubit frequency can be expanded up to the first order of $\Phi_n(t)$,
\begin{equation}
    \omega_q\big(\Phiext{\text{tot}}(t)\big) = \omega_q\big(\Phi(t)\big) + \left(\left. \frac{\partial \omega_q}{\partial \Phi}\right|_{\Phi(t)}\right)\Phi_n(t) + \mathcal{O}\left(\Phi_n(t)\right)^2.
\end{equation}
Consequently, the qubit $\ket{e}$ accumulates a phase $\varphi(t)$ relative to $\ket{g}$,
\begin{equation}\label{eqn:time-depedent_varphi}
\begin{aligned}
    \varphi(t) =& \int_0^{t} dt' \omega_q(\Phiext{\text{tot}}(t'))\\
            \approx& ~\tilde{\omega}_q t + \int_0^t dt'\left( \left.\frac{\partial \omega_q}{\partial \Phiext{}} \right|_{\Phiext{}(t')} \right) \left(\sum_i \Phi_i(\omega_i) \sin{(\omega_i t' + \xi_i)} \right),
\end{aligned}
\end{equation}
that depends on the derivative of the flux dispersion $\dot{\omega}_q = \partial \omega_q/\partial \Phiext{}$, where $\tilde{\omega}_q$ is the mean qubit frequency under the flux drive. 
As a result of the noisy $\xi_i$, the uncertainties in the qubit phase directly leads to the dephasing of the qubit according to the decay function~\cite{Ithier2005} $e^{-\braket{\delta \varphi^2(t)}/2}$, where $\braket{\cdot}$ is an average over statistical ensembles, or different realizations of the stochastic phase $\xi_i$.

For a constant flux bias, $\dot{\omega}_q = \alpha$ is independent of $t$, and the integration of \autoref{eqn:time-depedent_varphi} reduces to an uncertain phase,
\begin{equation}
    \delta \varphi(t) = \alpha t\sum_i \Phi_i(\omega_i) \frac{\sin{(\omega_i t/2)}}{\omega_i t/2}\sin{(\omega_i t/2 + \xi_i)}.
\end{equation}
Applying the identity $\braket{\sin^2{\xi_i}} = 1/2$ and replacing the summation over $\omega_i$ with an integration, we find the exponent of the decay function,
\begin{equation}\label{eqn:ramsey_filter}
    \frac{\braket{\delta \varphi^2(t)}}{2} =  \frac{t^2 \alpha^2}{2\pi} \int_{-\infty}^{\infty} d\omega S_{\Phi}(\omega) \text{sinc}^2 \left(\frac{\omega t}{2} \right).
\end{equation}
If the qubit is kept at the degeneracy position with $\alpha = 0$, the random phase $\xi_i$ does not contribute any noise to the qubit phase $\varphi(t)$, hence the protection against flux-noise-induced dephasing.
Away from $\Phi_0/2$, the nonzero flux-dispersion $\alpha$ translates uncertainties in flux to uncertainties in qubit frequency and ultimately leads to dephasing. In particular, the noise spectral density is weighted by a sinc function centered about $\omega = 0$. Because flux noises are predominantly $1/f$-like in their power spectral density, the larger susceptibility to low frequency flux noise is particularly detrimental to qubit dephasing.

\begin{figure}
    \centering
    \includegraphics{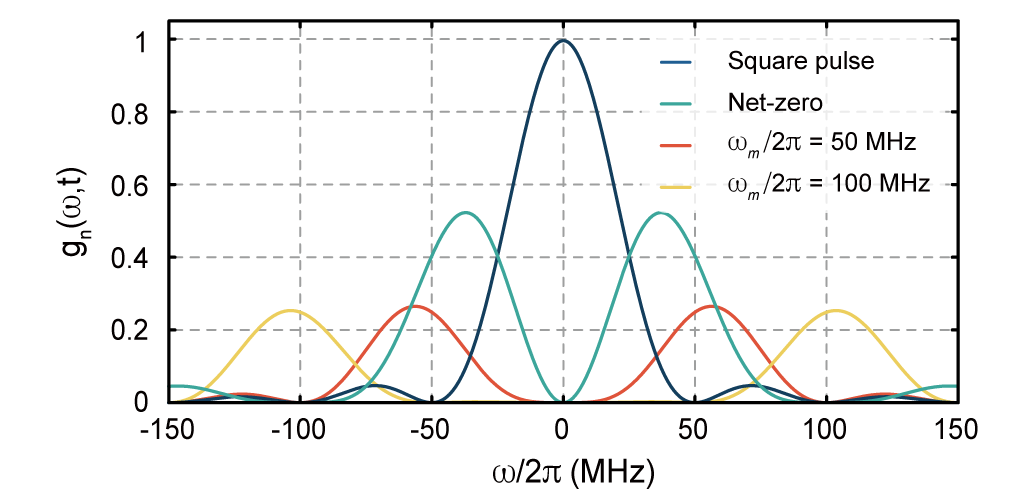}
    \caption{We plot the frequency domain noise-filtering function $g_n(\omega,t)$ for $t = 20~\text{ns}$. Compared to the net-zero pulse, whose filtering function we adapt from Ref.~\cite{Ithier2005}, a sinusoidal modulation with $\omega_m = 2\pi\times 50~\text{MHz}$ provides better noise filtering. Importantly, we can choose \textit{in situ} the position of the peaks in $g_n(\omega,t)$ by altering the sinusoidal modulation frequency $\omega_m$.}    
    \label{fig:DD_filter}
\end{figure}

By shifting the center of the weighting function to the modulation frequency $\omega_m$, our sinusoidal dynamical decoupling scheme therefore could significantly reduce the qubit dephasing rate. As we discussed in the main text, a sinusoidal flux modulation about the degeneracy position, $\Phi(t) = \Phi_0/2 + \delta\Phi \sin{(\omega_m t)}$, causes a near sinusoidal modulation on the derivative of the flux dispersion, $\dot{\omega}_q(t) \approx \alpha \sin{(\omega_m t)}$. Inserting this expression into \autoref{eqn:time-depedent_varphi} and following the same derivation, we arrive at an approximate expression for qubit dephasing under our dynamical decoupling scheme,
\begin{equation}\label{eqn:sinosoidal_filter}
\begin{aligned}
    \frac{\braket{\delta \varphi^2(t)}}{2} \approx \frac{t^2 \alpha^2}{8\pi} \int_{-\infty}^{\infty} d\omega S_{\Phi}(\omega) \bigg\{ \text{sinc}^2 &\left[\frac{(\omega+\omega_m)t}{2}\right] + \text{sinc}^2 \left[\frac{(\omega-\omega_m)t}{2}\right] 
    \\& + 2\text{sinc} \left[\frac{(\omega-\omega_m)t}{2}\right]\text{sinc} \left[\frac{(\omega+\omega_m)t}{2}\right]\cos{(\omega_m t)}\bigg\},
\end{aligned}
\end{equation}
where the approximation comes from treating $\dot{\omega}_q(t)$ as a sinusoidal function. Consequently, we could choose which part of the noise spectral density to sample by choosing the modulation frequency $\omega_m$.
\autoref{fig:DD_filter} further illustrates this filtering effect by explicitly plotting the frequency domain noise-filtering function $g_n(\omega,t)$ defined according to~\cite{Bylander2011},
\begin{equation}
    \frac{\braket{\delta \varphi^2(t)}}{2} = \frac{t^2 \alpha^2}{2\pi} \int_{-\infty}^{\infty} d\omega S_{\Phi}(\omega) g_n(\omega,t),
\end{equation}
with a free evolution time $t = 20~\text{ns}$ equal to our CZ gate duration.

Finally, we note that because of their different $S_{\Phi}(\omega)$, different sources of flux noise are affected differently by the dynamical decoupling scheme. In this work, we are mostly concerned about two types of the error sources: those that behave like white-noise, and those that behave like $1/f$ noise,
\begin{equation}
    \braket{\delta \varphi^2(t)} = \braket{\delta \varphi^2(t)}_\text{white} + \braket{\delta \varphi^2(t)}_{1/f}.
\end{equation}
While our dynamical decoupling scheme can greatly reduce the qubit dephasing due to the $1/f$-noise via frequency selection, it has little effect on that from the white noise.

\subsection{Comparing qubit coherence times under dynamical decoupling}\label{sec:decoupling_experiment}

\begin{figure}[!]
    \centering
    \includegraphics{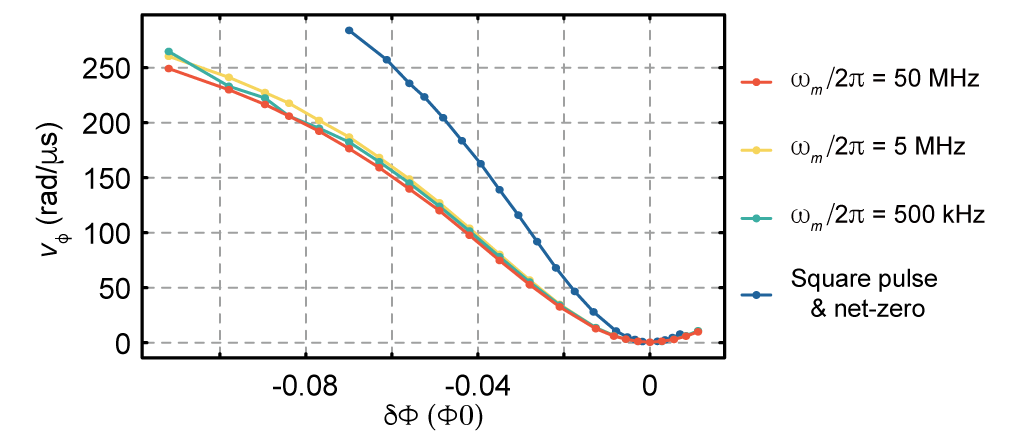}
    \caption{We measure the accumulated conditional phase when both qubits are subjected to the same flux modulation. Averaged over integer periods of the modulation's temporal envelope, the accumulation speed $v_\phi$ is plotted as a function of the modulation amplitude $\delta \Phi$.}
    \label{fig:z2vPhi}
\end{figure}

To demonstrate the efficacy of our dynamical decoupling scheme, we aim to compare the qubit coherence times under different flux control pulses $\delta \Phi(t)$. However, a direct comparison at the same driving amplitude across different pulse shapes does not provide much insight in terms of gate-operations because they would, in general, lead to different CZ gate duration. Indeed, the pulse shape of the flux drive relates to the temporal envelope of the \textit{ZZ}-interaction strength $g_{zz}(t) \propto \dot{\omega}_q^A(t) \dot{\omega}_q^B(t)$ through the time-dependence of $\dot{\omega}(t)$. \autoref{fig:z2vPhi} shows the measured accumulation speed of the conditional phase, $v_\phi$, averaged over integer periods when both qubits are subjected to the same flux drive of amplitude $\delta \Phi$. Compared to square modulations of the same amplitude, a sinusoidal flux modulation applied to both qubits would require almost twice the duration to perform a CZ gate. 

\begin{figure}[!]
    \centering
    \includegraphics{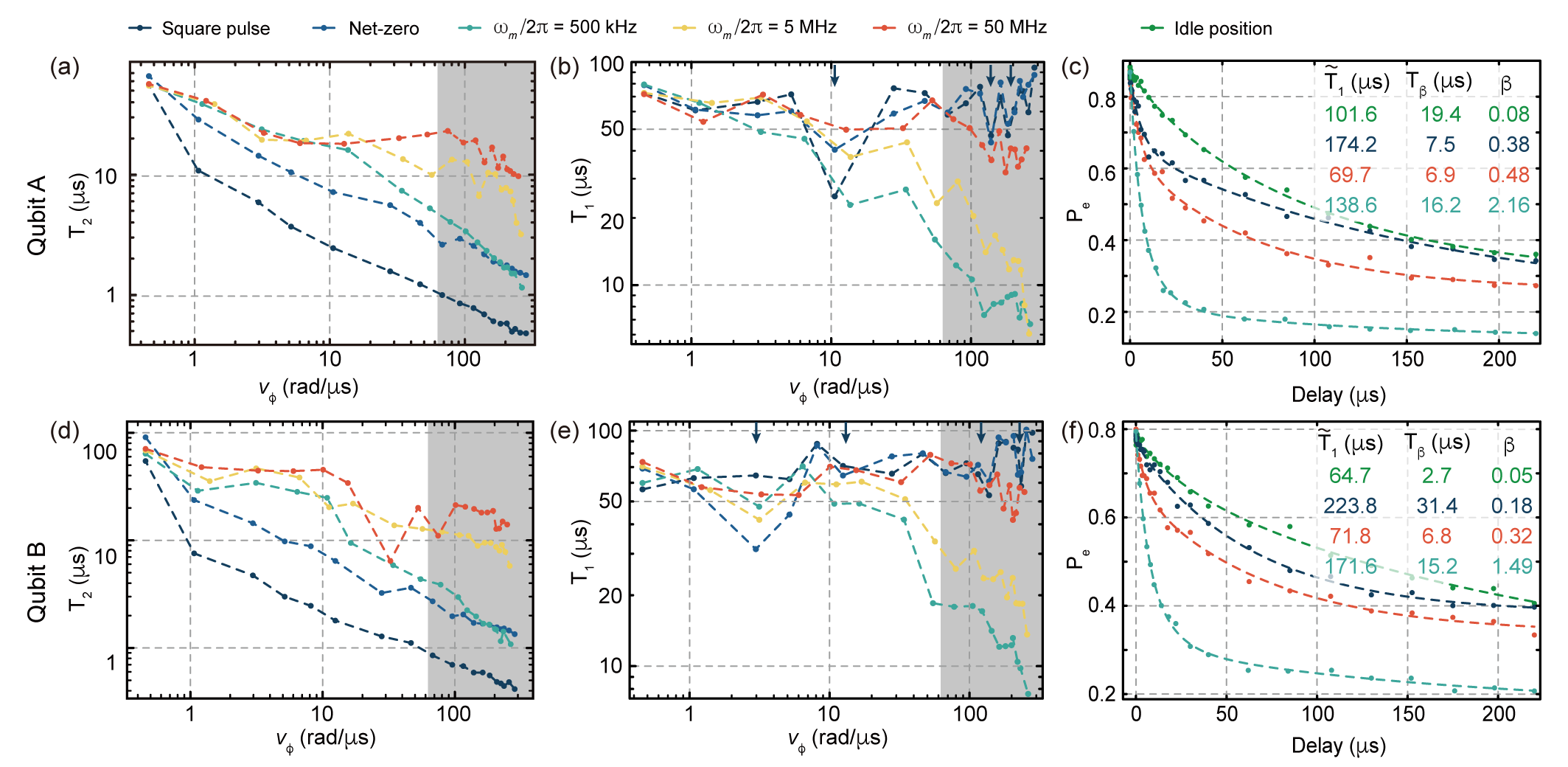}
    \caption{Using different modulation pulse shapes, we keep qubit B at $\Phi_0/2$ and measure, as a function of the modulation amplitude $\delta \Phi$, qubit A's (a) Ramsey characteristic times $T_2$ (identical to Fig.~3(b) of the main text) and (b) characteristic depolarization times $T_1$, both corresponding to the time it takes for the measured decay functions to be reduced to $1/e$.
    $\delta \Phi$ is then converted to $v_\varphi$ had the same flux modulation been applied to both qubits according to \autoref{fig:z2vPhi}.
    Shaded regions correspond to $v_\phi$ large enough to support CZ gates under $50~\text{ns}$. 
    Vertical downward arrows in (b) identifies the TLS that exchange energy with qubit A under static (square wave and net-zero pulse) flux biases and cause fast qubit depolarization. The effects of these TLS are better illustrated by the double-exponential decay of individual qubit decay curves. 
    (c) We show characteristic decay curves measured close to the CZ operation position ($v_\phi \approx 157~\text{rad}/\mu\text{s}$) when $\Phi_A(t)$ is either statically biased (dark blue) or under sinusoidal modulations (red and cyan). For comparison, we also plot the decay curve when the qubit A is kept at its idle position (green). To account for the double-exponential decay, we fit these data using $\braket{P_e(t)} \propto e^{\beta \big(\text{exp}(-t/T_\beta)-1\big)}e^{-t/\tilde{T}_1}$~\cite{pop2014,Gustavsson2016}, where a higher value of $\beta$ reflects a larger damage done by the TLS.
    In (d,e,f), the same experiments are repeated on qubit B. However, because qubit B has a smaller frequency at its degeneracy position $\omega_B^0 < \omega_A^0$, keeping qubit A at the degeneracy position would risk decohering qubit B through an iSWAP-like interaction with qubit A when $\Phiext{B}$ is modulated. In our CZ gate, this decoherence channel is eliminated by ensuring the adiabatic limit (see Sec.~\ref{sec:XX} and \ref{sec:LandauZener}). Here, we eliminate this decoherence channel by keeping qubit A at $\Phiext{A} = 0$, where the native $g_{zz}$ is also 0 (see \autoref{fig:coupling_landscape}).}
    \label{fig:coherenceVmodulation}
\end{figure}

Instead, we compare the qubit coherence times across different pulse shapes while keeping constant the CZ gate duration, or $v_\phi$, had both qubits been driven by the same flux pulse. For each qubit, we modulate its external flux using different amplitudes $\delta \Phi$ and pulse shapes, and measure its depolarization time $T_1$ and Ramsey decay profile $\tilde{P}_e(t)$ while the other qubit is kept at a place where the native \textit{ZZ}-interaction is turned off. In theory, the Ramsey decay profile~\cite{Ithier2005}
\begin{equation}
    \tilde{P}_e(t) = \tilde{P}_e(0)\cdot \text{exp}\left(- \frac{t}{2T_1}\right)\cdot\text{exp}\left(-\frac{1}{2}\braket{\delta \varphi^2(t)}_\text{white}\right)\cdot\text{exp}\left(-\frac{1}{2}\braket{\delta \varphi^2(t)}_{1/f}\right)
\end{equation}
allows for a complete separation of different dephasing sources and their effects. However, to avoid over-fitting our data and introducing bias, we simply report the characteristic time $T_2$ for the measured Ramsey decay profile to reduce to $1/e$. Because $\braket{\delta \varphi^2(t)}_\text{white}$ mostly likely arises from photon shot-noise or state measurement/preparation errors, it should be largely independent of the external flux applied to the qubits. Changes in $T_2$ therefore reflect changes in $\braket{\delta \varphi^2(t)}_{1/f}$ and $T_1$.
\autoref{fig:coherenceVmodulation} shows the extracted $T_1$ and $T_2$ with $\delta \Phi$ converted into $v_\phi$ according to \autoref{fig:z2vPhi}. 

Additionally, \autoref{fig:coherenceVmodulation} reveals a non-negligible dependence of the qubits' $T_1$ on the applied external flux. When the qubit is statically (square wave and net-zero pulses) biased away from the degeneracy position, a slight improvement in $T_1$ can be observed as a result of disjoint support~\cite{Huang2021}, arising from increasingly localized qubit states in phase space. Ref.~\cite{Huang2021} also predicts an increase of $T_1$ for the sinusoidally modulated qubits compared to those parked at the degeneracy position, which we do not observe. This is likely because the qubit crosses many TLS (downward arrows in \autoref{fig:coherenceVmodulation}(b,e)) while being flux modulated. Indeed, $T_1$ measurements (\autoref{fig:coherenceVmodulation}(c,f)) performed near our CZ operation position ($v_\phi\sim 157~\text{rad}/\mu\text{s}$) show clear signs of double-exponential decay, symptomatic of qubits afflicted by TLS poisoning.
Consequently, slow modulations with $\omega_m = 2\pi\times 500~\text{kHz}$ and $\omega_m = 2\pi\times5~\text{MHz}$ result in a higher probability for the qubit to depolarize via an adiabatic exchange of energy with the TLS. With a fast modulation frequency $\omega_m = 2\pi\times50~\text{MHz}$, however, the qubit $T_1$ remains roughly constant across all modulation amplitudes.

The effectiveness of our dynamical decoupling in combating the $1/f$-noise is therefore clearly demonstrated in the significant improvements in $T_2$ despite the reductions in $T_1$.

\section{Two-qubit CZ gate}

Embedding the sinusoidal dynamical coupling scheme into our flux control sequences, we calibrate two-qubit CZ gates. Because the CZ operation relies on an adiabatic process that alters $g_{zz}$ while preserving the qubit excitation, we first derive in Sec.~\ref{sec:LandauZener} the adiabatic limit that bonds $\omega_m$.
In Sec.~\ref{sec:gateCal}, we detail the pulse-level calibration sequences for the CZ gate. We also identify large distortions at the end of each flux control pulse that require us to append an idle time of $20~\text{ns}$ to every flux pulse we apply to the qubits.
Finally, in Sec.~\ref{sec:gate_error}, we discuss the error sources that limit our CZ gate fidelity. 

\subsection{The adiabatic limit}\label{sec:LandauZener}

We start by recalling from the main text that each fluxonium qubit is well described by a pair of persistent current states coupled through a tunneling energy $\Delta$. Written in the basis of these persistent current states ($\hatSigmaPrime{z}{i}$), the Hamiltonian of our coupled fluxonium system is,
\begin{equation}\label{eqn:Landau-Zener Hamiltonian}
\begin{aligned}
    H =& \sum_{i = A,B} \bigg(H_z^i \hatSigmaPrime{z}{i} + H_x^i\hatSigmaPrime{x}{i} \bigg) +  J\hatSigmaPrime{z}{A}\hatSigmaPrime{z}{B}\\
    =&  \frac{\hbar}{2}\sum_{i = A,B} \bigg[ \left(\epsilon_i + \frac{J}{\hbar}\hatSigmaPrime{z}{j}\right) \hatSigmaPrime{z}{i} + \Delta_i\hatSigmaPrime{x}{i} \bigg]
\end{aligned}
\end{equation}
where the interaction term is absorbed into $H_z^i$ of each qubit, $\epsilon_i = I_p^i \left(\Phiext{i} - \Phi_0/2 \right)$ is the magnetic energy difference between the two persistent current states of qubit $i$, and the effect of the flux modulation conveniently enters through its time dependence $\epsilon_i(t)$. 

For each qubit, the non-adiabatic errors are typically understood using the Landau-Zener transitions~\cite{Landau1932,Zener1932}, where a linear trajectory $H_z(t) = \dot{\epsilon} t$ going from large negative time to large positive time induces a probability of exciting the qubit~\cite{Martinis2014},
\begin{equation}\label{eqn:landau-zener probability}
    P_e = \text{exp}\left(- \frac{\pi \Delta^2}{\hbar \dot{\epsilon}} \right).
\end{equation}
A small error of $P_e = 10^{-4}$ implies the condition $\hbar \dot{\epsilon} = 0.341 \Delta^2$. In this work, we modulate the external flux of either qubit within $0.5 \pm 0.15 \Phi_0$, which, according to our qubit parameters in \autoref{tab:device_params}, corresponds to a maximum deviation of $\pm \delta \epsilon_\text{max} \approx 2\pi \times 1.9 ~\text{GHz}$ in both qubits. The above condition therefore is satisfied when the external flux is altered by $\delta \Phi = 0.15 \Phi_0$ over a duration longer than $1.2~\text{ns}$ for qubit A or $2.1~\text{ns}$ for qubit B, comparable to our choice of $2~\text{ns}$ rise time in Fig.~2(c) of the main text.
In the case of a sinusoidal modulation with $\epsilon(t) = \delta \epsilon \sin{(\omega_m t)}$, we could not directly invoke \autoref{eqn:landau-zener probability}, but instead use the slope of $\epsilon(t)$ to bound $\dot{\epsilon} \leq \omega_m \delta \epsilon_\text{max}$. Again requiring the non-adiabatic error to be smaller than $10^{-4}$, we find a maximum modulation frequency of $2\pi\times 133~\text{MHz}$ for qubit A and $2\pi\times 76~\text{MHz}$ for qubit B. For our CZ gates, we choose $\omega_m = 2\pi\times 50~\text{MHz}$, well below these thresholds.

We stress that the errors estimated above accounts for all non-adiabatic errors, including those arising from the \textit{XX}- and \textit{XZ}-interactions. Indeed, the native \textit{XX}- and \textit{XZ}- interactions discussed in Sec~\ref{sec:XZ} and \ref{sec:XX} both emerge from the coupling term $J\hatSigmaPrime{z}{A}\hatSigmaPrime{z}{B}$ in \autoref{eqn:Landau-Zener Hamiltonian} by simple rotations to the instantaneous energy eigenbasis. Therefore, we do not perform additional calibrations or corrections on these unwanted interactions, but instead content that our two-qubit entangling operations conserve the individual qubit occupation up to a negligible non-adiabatic error below $10^{-4}$. 

\subsection{Tuning up a CZ gate}\label{sec:gateCal}

In the absence of non-adiabatic qubit rotations, the entangling interaction realized by our native \textit{ZZ}-interaction is described by a diagonal unitary transformation,
\begin{equation}
    \hat{U}_\text{ZZ} = \begin{pmatrix}
    1 & 0 & 0& 0\\
    0 & e^{i \zeta_A} & 0 & 0\\
    0 & 0 & e^{i \zeta_B} & 0\\
    0 & 0& 0 & e^{i(\zeta_A+\zeta_B + \phi)}
    \end{pmatrix},
\end{equation}
that only differs by single qubit phases $\zeta_{A,B}$ from a CZ gate, which has the two-qubit conditional phase $\phi = \pi$. In this section, we detail the procedures used to calibrate these phases. 

To start, we calibrate our flux control lines following the techniques outlined in Ref.~\cite{Bao2022}. In addition, we fine-tune the crosstalk between these lines using the periodicity of the native \textit{ZZ}-interaction in external flux (see \autoref{fig:coupling_landscape}). Specifically, with qubit $i$ biased away from its flux degeneracy position, we measure and ensure its resonant frequency remains constant when the other qubit $j$ is biased at integer multiples of $\Phi_0/2$.

To realize a CZ gate, we simultaneously modulate the external flux of both qubits with,
\begin{equation}
    \Phiext{A,B}(t) = \Phi_0/2 - \delta \Phiext{A,B}\sin{\left(\omega_m t \right)},
\end{equation}
for a duration of $t_\text{CZ}$ such that by the end of the modulation, the conditional phase accumulates to $\phi = \pi$.
In this work, we choose a modulation frequency $\omega_m = 2\pi\times 50~\text{MHz}$. Because the sinusoidal dynamical decoupling scheme is only effective for gate durations that are integer multiples of the modulation period, we then choose $t_\text{CZ} = 20~\text{ns}$. 

\begin{figure}[!]
    \centering
    \includegraphics{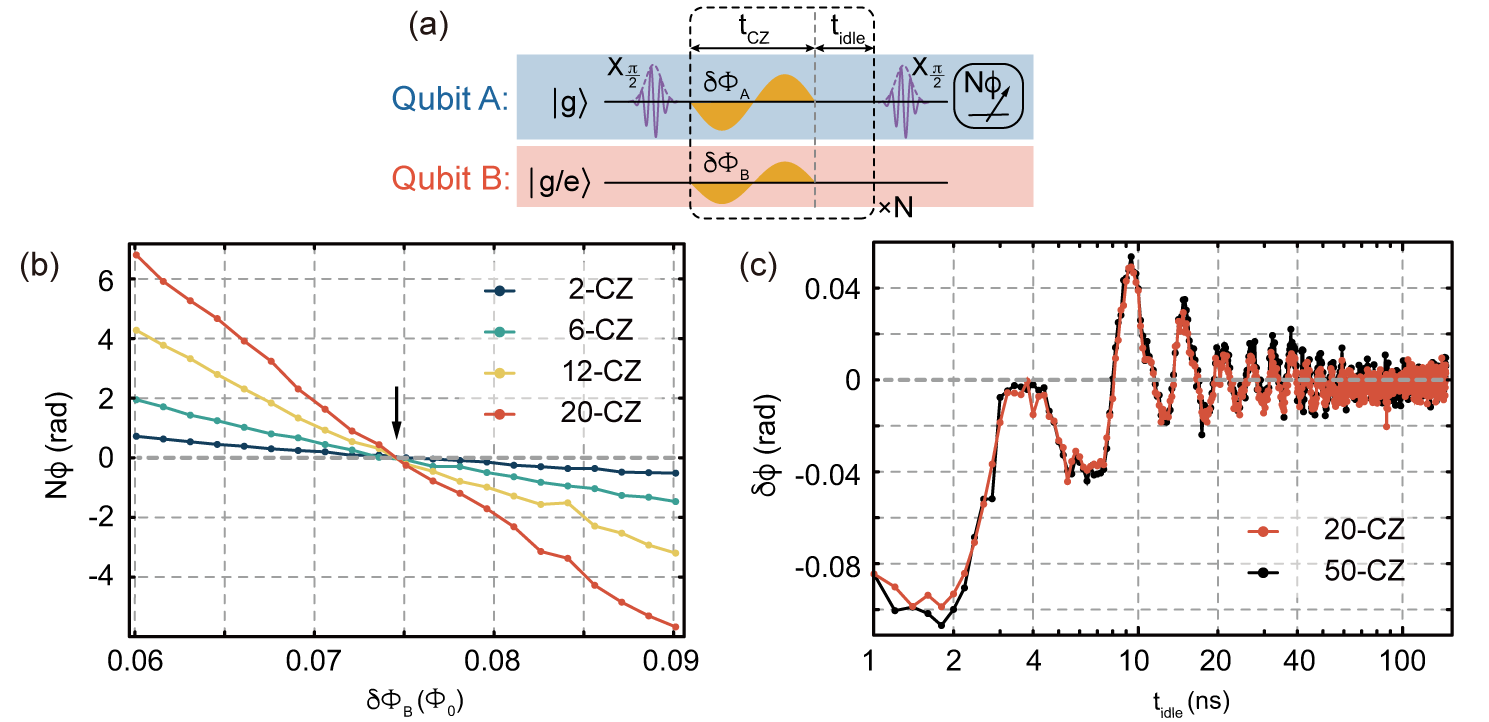}
    \caption{(a) For a given modulation amplitude $\delta \Phiext{A}$ on qubit A, we calibrate a CZ gate by varying $\delta \Phiext{B}$ applied on qubit B and measuring the accumulated conditional phase $N\phi$ on qubit A after $N$ consecutive operations. $\delta \Phiext{B}$ can be accurately determined by the location where $N\phi = 0$ for all even $N$.
    (b) As an example, we show the measured $N\phi$ as a function of $\delta \Phiext{B}$ with $\delta \Phiext{A} = 0.12 \Phi_0$, $\omega_m = 2\pi\times50~\text{MHz}$, $t_\text{CZ} = 20~\text{ns}$, and $t_\text{idle} = 20~\text{ns}$. Under these conditions, $\delta \Phiext{B} = 0.0745 \Phi_0$ (vertical black arrow) leads to an accurate conditional phase $\phi = \pi$.
    (c) Using the pulse sequence in (a), we can also explicitly measure the flux-distortion-induced deviations in the conditional phase $\delta \phi$ as a function of the idle time $t_\text{idle}$. After removing the effects of $g_{zz}^\text{res}$ during $t_\text{idle}$, we reference $\delta \phi$ to the average conditional phase measured between $t_\text{idle} = 100$ and $150~\text{ns}$.
    }
    \label{fig:cz_calibration}
\end{figure}

For a given modulation amplitude $\delta\Phiext{A}$ on qubit A, we calibrate the necessary modulation amplitude $\delta\Phiext{B}$ applied on qubit B to ensure $\phi = \pi$. As shown in \autoref{fig:cz_calibration}(a,b), for each value of $\delta \Phiext{B}$, we amplify the error on $\phi$ by applying a series of $N$-CZ operations. Because each CZ-gate contributes a conditional phase of $\pi$, an even number of CZ-gate should always return the conditional phase to $0$, thereby allowing for an accurate determination of $\delta \Phiext{B}$.

In this work, we had to append an idle time of $t_\text{idle} = 20~\text{ns}$ after each CZ gate to combat the detrimental effects from a relatively large flux-pulse distortion. In \autoref{fig:cz_calibration}(c), we explicitly characterize this detrimental effect by measuring the deviations of the conditional phase $\delta \phi$ from $\phi = \pi$ as a function of the idle time $t_\text{idle}$.
In this measurement, we subtract from $\phi$ the extra conditional phase arising from $g_{zz}^\text{res}$ during $t_\text{idle}$ (see Sec.~\ref{sec:XX}), and the resulting $\phi(t_\text{idle})$ remains constant for $t_\text{idle} > 100~\text{ns}$. A reference phase $\tilde{\phi}$, equal to the statistical mean of $\phi(t_\text{idle})$ measured within $t_\text{idle} \in [100,150]~\text{ns}$, is then subtracted to find $\delta \phi (t_\text{idle}) = \phi (t_\text{idle}) - \tilde{\phi}$.
While $\delta \phi$ may appear deterministic in our experiment with constant separations between adjacent CZ operations, it is transformed to an uncertainty in $\phi$ in a realistic quantum circuit, where adjacent CZ operations are interrupted by an unknown number of single-qubit operations.
By choosing $t_\text{idle} = 20~\text{ns}$, we therefore sacrifice the overall gate speed in favor of an improved control over the conditional phase with $\delta \phi \sim 0.02~\text{rad}$. According to \cite{Ghosh2013}, we predict this uncertainty in $\phi$ corresponds to a CZ-gate infidelity of $2\times 10^{-5}$.

The last step to calibrating a CZ gate for any given $\delta \Phiext{A}$ is to 
correct the single-qubit phases $\zeta_{A,B}$ accumulated during the CZ operation using virtual-Z gates~\cite{McKay2017}. For a rough calibration, we can extract $\zeta_{A(B)}$ by measuring the single-qubit phase accumulated on qubit A(B) versus the number of CZ operations while the other qubit is prepared in $\ket{g}$. Fine-tuning of the virtual-Z gates is achieved by numerically optimizing~\cite{Kelly2014} the XEB sequence fidelity~\cite{Neill2018,Arute2019} at a fixed length of $m = 50$.

\begin{figure}
    \centering
    \includegraphics{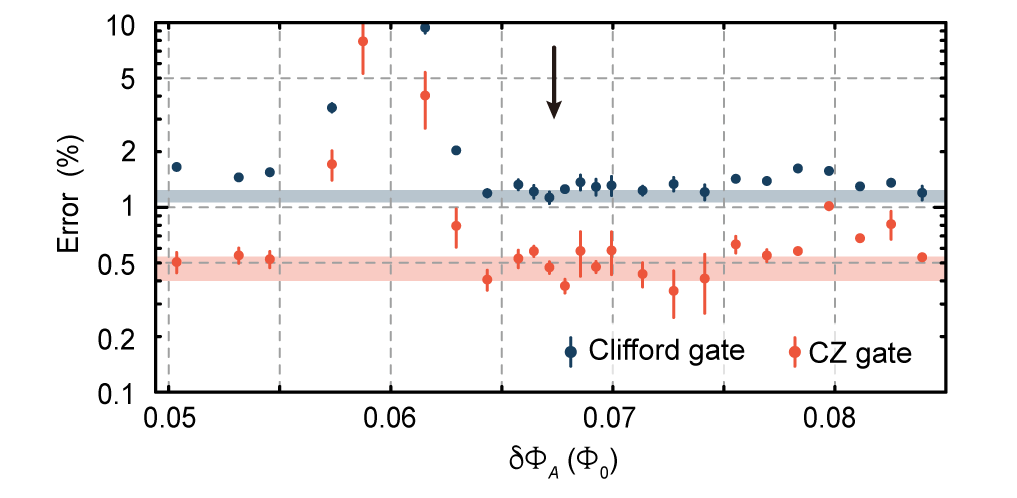}
    \caption{For each $\delta \Phiext{A}$, we calibrate a CZ gate and then measure its error (red), as well as the error of the Clifford gate (blue), using standard RB. Each point corresponds to the mean fidelity measured from at least two separate RB experiments, and the error bars correspond to the standard deviation of the errors in time. Vertical black arrow points to the $\delta \Phiext{A}$ we used for Fig.~4 of the main text, and the shaded regions correspond to the one standard deviation about the mean errors reported in the main text.
    }
    \label{fig:optimzing_PhiA}
\end{figure}

Finally, we explore the parameter space of $\delta \Phiext{A}$ to find the optimal operating position. For each $\delta \Phiext{A}$, we calibrate a CZ gate following the procedures outlined above and characterize it using standard RB techniques~\cite{Magesan2012,Barends2014}.
Because our qubit system suffers from fluctuating coherence times (see Sec.~\ref{sec:general} and \ref{sec:SQ-gate}), we perform at least two sets of RB measurements at each $\delta \Phiext{A}$ to gain a rough sense on the fidelity's fluctuation in time. 
\autoref{fig:optimzing_PhiA} shows, as a function of $\delta \Phiext{A}$, the mean error (and its uncertainty) on the CZ gates and on the corresponding Clifford gates. 
Likely caused by significant energy exchanges with some TLS close in frequency to the qubit, we find the CZ gates to perform particularly poorly around $\delta \Phiext{A} = 0.06~\Phi_0$. However, away from this position and up to fluctuations in time, we find the gate errors to remain largely constant between $\delta \Phiext{A} = 0.065~\Phi_0$ and $0.075~\Phi_0$, demonstrating both the robustness of our gate and the robustness of our calibration procedure. In Fig.~4 of the main text, we chose to perform the CZ gates at $\delta \Phiext{A} = 0.0673 ~\Phi_0$ (black vertical arrow).
For reference, the shaded regions correspond to the fidelities' one standard deviation about the mean reported in the main text.

\subsection{CZ gate error analysis}~\label{sec:gate_error}

We start our error analysis on the two-qubit gate errors with a consistency check among the measured gate fidelities. We follow the same decomposition scheme of Ref.~\cite{Barends2014}, where each two-qubit Clifford gate contains on average $\frac{33}{4}$ single-qubit primary gates and $\frac{3}{2}$ CZ gates. Correspondingly, the error rate of a Clifford gate $r_C$ is given by
\begin{equation}
    r_C = \frac{33}{4}r_\text{SQ} + \frac{3}{2}r_\text{CZ},
\end{equation}
where $r_\text{SQ}$ is the average single-qubit gate error and $r_\text{CZ}$ is the CZ gate error. As shown in the main text and in Sec.~\ref{sec:SQ-gate}, $r_\text{SQ} = 0.05(6) \pm 0.00(4)\%$ and $r_\text{CZ} = 0.47 \pm 0.07 \%$. These measured error rates correspond to a predicted Clifford error rate $r_C \approx 1.53 \pm 0.12\%$, roughly consistent with the measured Clifford error of $r_C^m = 1.15 \pm 0.08\%$.

The CZ gate error $r_\text{CZ}$ can be further decomposed into decoherence errors and control errors. The nonadiabatic errors discussed in Sec.~\ref{sec:LandauZener} contribute approximately $0.01 \%$ per qubit to control errors. Additionally, the uncertainty in the conditional phase (Sec.~\ref{sec:gateCal}) leads to a control error of approximately $0.002\%$.

We attribute the majority of the remaining error $r^m \approx 0.45 \pm 0.07\%$ to decoherence errors.
The decoherence limit of a two-qubit gate can in general be estimated using~\cite{OMalley2015},
\begin{equation}\label{eqn:decoherence_limit}
    r = \frac{1}{6}\sum_{i = A,B}\bigg\{
     \frac{2t_\text{idle}}{T_{1,\text{idle}}^i} + \braket{\delta \varphi_i^2(t_\text{idle})}_\text{white} + 
    \frac{2t_\text{CZ}}{T_{1,\text{CZ}}^i} + \braket{\delta \varphi_i^2(t_\text{CZ})}_\text{white} + \braket{\delta \varphi_i^2(t_\text{CZ})}_{1/f}
    \bigg\},
\end{equation}
where the decoherence of the qubit $i$ at its idle position and CZ-operation position are respectively parameterized by its energy decay time $T_{1,\text{CZ}}^i$ and $T_{1,\text{idle}}^i$, and dephasing effects $\braket{\delta \varphi_i^2(t_\text{CZ})}$ and $\braket{\delta \varphi_i^2(t_\text{idle})}$ for both white- and $1/f$-noise sources.
Because $\braket{\delta \varphi^2(t)}_\text{white}$ mostly likely arises from photon shot-noise, it should be largely independent of the external flux applied to the qubits, $\braket{\delta \varphi^2(t_\text{idle})}_\text{white} \approx \braket{\delta \varphi^2(t_\text{CZ})}_\text{white}$. Additionally, by assuming $T_{1,\text{CZ}}^i \approx T_{1,\text{idle}}^i$, we can approximately determine the decoherence limit due to the first four terms of \autoref{eqn:decoherence_limit} experimentally by measuring the fidelity of an idle gate of duration $t = t_\text{CZ} + t_\text{idle} = 40~\text{ns}$, $\overline{{F}_\text{idle}} = 99.78 \pm 0.05\%$. Based on the measured idle fidelities of the single-qubit gates (Sec.~\ref{sec:SQ-gate}), we predict a decoherence limit for a $40~\text{ns}$ idle gate $r_\text{idle} = 0.18 \pm 0.07\%$, consistent with the measured $r_\text{idle}^m = 1-\overline{{F}_\text{idle}} =  0.22\pm 0.05\%$. As we mentioned in the main text, half of the total CZ gate error can be account for by this decoherence error from the idle position. We can also estimate the effect of the last term in
\autoref{eqn:decoherence_limit} using our measured $1/f$-noise spectrum $S_{\Phi}(f) = A^2/2|f|$ with $A \approx 10.6~\mu\Phi_0/\sqrt{\text{Hz}}$ for both qubits and \autoref{eqn:sinosoidal_filter} to find a negligible $r_{1/f} \approx 0.002\%$. In contrast, \autoref{eqn:ramsey_filter} predicts an order-of-magnitude larger decoherence error $r_{1/f}^\prime \approx 0.024\%$ when no dynamical decoupling scheme is employed.
Finally, we note that the decoherence effect from the large number of TLS the qubits sample along their flux modulation path have not been accounted for. Indeed, as discussed in Sec.~\ref{sec:decoupling_experiment}, these TLS could lead to rapid qubit depolarization, and significantly impact the CZ gate fidelity. We therefore attribute the remaining errors to their effect.

\bibliographystyle{apsrev4-2}
\bibliography{ref}